\input epsf
\documentclass[nohyper,notoc]{article} 
\usepackage{axodraw}
\usepackage{epsfig}

\def\bit{\begin{itemize}}
\def\eit{\end{itemize}}
\def\ben{\begin{enumerate}}
\def\een{\end{enumerate}}
\def\beq{\begin{equation}}
\def\eeq{\end{equation}}
\def\bea{\begin{eqnarray}}
\def\eea{\end{eqnarray}}
\def\bq{\begin{quote}}
\def\eq{\end{quote}}
\def \lsim{\mathrel{\vcenter
     {\hbox{$<$}\nointerlineskip\hbox{$\sim$}}}}
\def \gsim{\mathrel{\vcenter
     {\hbox{$>$}\nointerlineskip\hbox{$\sim$}}}}
\def\gappeq{\mathrel{\rlap {\raise.5ex\hbox{$>$}}
{\lower.5ex\hbox{$\sim$}}}}
\def\lappeq{\mathrel{\rlap{\raise.5ex\hbox{$<$}}
{\lower.5ex\hbox{$\sim$}}}}

\def\m3e{\mu \to e \bar{e} e}

\def\m{\mu}

\def\hatt{\hat{t}}
\def\hats{\hat{s}}
\def\hatu{\hat{u}}

\begin{document}
%\vspace*{-1in}
\renewcommand{\thefootnote}{\fnsymbol{footnote}}
\begin{center}
{\Large {\bf 
Constraining flavoured contact interactions at the LHC}}
\vskip 25pt
{\bf   
Sacha Davidson $^{1,}$
\footnote{s.davidson@ipnl.in2p3.fr} 
and  S\'ebastien Descotes-Genon $^{2,}$
\footnote{descotes@th.u-psud.fr}  }
 
\vskip 10pt  
$^1${\it IPNL, Universit\'e de Lyon, Universit\'e Lyon 1, CNRS/IN2P3, 4 rue E. Fermi 69622, Villeurbanne cedex, France
}\\
$^2${\it Laboratoire de Physique Th\'eorique,
CNRS/Univ. Paris-Sud 11 (UM8627), 91405 Orsay Cedex.
 } \\
\vskip 20pt
{\bf Abstract}
\end{center}

\begin{quotation}
  {\noindent\small 
Contact interactions are the low-energy footprints of New Physics, so
ideally, constraints upon them should be as generic and 
model independent as possible.  Hadron colliders 
search for four-quark contact
interactions with incident valence quarks, and the LHC currently 
sets limits on a flavour sum (over $uu,dd$ and $ud$) of 
selected interactions. We  approximately translate
these bounds to a more complete (and larger) set of dimension-six 
interactions of specified flavours. These estimates are
obtained
at the parton level,  are mostly analytic and are less restrictive than the 
experimental bounds on flavour-summed interactions.  
The estimates may scale in a simple way to higher energy
and luminosity.
\vskip 10pt
\noindent
}

\end{quotation}

\vskip 20pt  

\setcounter{footnote}{0}
\renewcommand{\thefootnote}{\arabic{footnote}}

%%%%%%%%%%%%%%%%%%%%%%%%%%%%%%%%%%%%%%%%%%%%%%%%%%%%%%%%%%%%%%%%%%%%%%
%% INTRO        %%%%%%%%%%%%%%%%%%%%%%%%%%%%%%%%%%%%%%%%%%%%%%%%%%%%%%
%%%%%%%%%%%%%%%%%%%%%%%%%%%%%%%%%%%%%%%%%%%%%%%%%%%%%%%%%%%%%%%%%%%%%%
%\newpage
\section{ Introduction}
\label{intro}

Contact interactions inevitably arise as the low-energy remnants of high-energy theories. Should the LHC not  find (additional)  new particles, it can 
nonetheless be sensitive to
their traces in contact interactions. 
This paper focuses on four-quark contact interactions, with two 
incoming valence quarks, as is most probable at the LHC.
 New Physics from beyond  the 
LHC energy can induce various   operators~\cite{BW,polonais}, 
with  different Lorentz and gauge structures as well as 
flavour indices.
Bounds on contact interactions involving specific
flavours are  appropriate for constraining New
Physics models. The great variety of  high-energy  models
have different low-energy footprints, 
such as different flavour structures for contact interactions.
For instance, the contact interactions involving 
singlet (i.e., right-handed) $u$ quarks  might  differ from
those involving  $d$ quarks. 
Furthermore, the flavour structure of contact
interactions induced by New Physics is explored by precision flavour
physics (mostly for flavour off-diagonal operators),  so  the collider bounds should also
be for specific flavour indices, to allow
comparison and combination with  
low-energy observations.

However, the current LHC bounds are given for a subset
of  flavour-summed operators~\cite{ATLAS,CMS1,CMS2}.
The aim of this paper
is to make an approximate translation of the  collider
constraints onto individual, flavoured operators {and to illustrate the limits of such a translation}.
Two issues arise in attempting to apply the published 
contact interactions bounds to a different operator:
the Lorentz and  gauge  structure of the
operator  affects the partonic cross section, and 
the flavour indices 
affect possible interferences
with QCD, as well as controlling the
probability of finding the initial state
quarks in the proton. Let us mention that
four-quark contact interactions have been searched for
at the Tevatron in $q\bar{q} \to $ dijets~\cite{D0}, 
and at the LHC in $qq\to $ dijets~\cite{ATLAS,CMS1,CMS2}.
In both cases, the initial state $q$ or $\bar{q}$
are first generation valence quarks, whereas the final state
can be of any flavour, so the Tevatron and LHC can constrain
different flavour structures. In this paper, we focus on
the $qq \to$ dijets process at the LHC, 
recently used to constrain models with 
compositeness and/or heavy coloured octets
\cite{VS,colourons}.

Section \ref{sec:cin}  reviews the 
kinematical variables used by the experimental collaborations~\cite{ATLAS,CMS1,D0} to
constrain contact interactions from the rapidity distribution
of high energy dijets. These variables 
allow an approximate ``factorisation'' of  the
$pp \to$ dijets cross section into an integral
over parton distribution functions (pdfs), multiplying a partonic cross section.
In section~\ref{sec:op}, a basis of
Standard Model gauge-invariant, dimension-six effective
operators are listed, which, in  the presence of
electroweak symmetry breaking, 
induce the effective interactions
listed in section~\ref{sec:inteff}.
To obtain bounds as widely applicable as possible,
 we will constrain the coefficients of these effective
interactions
(this is discussed in section~\ref{sec:opinteff}).
{Issues regarding} flavour and interferences are discussed in section
\ref{sec:sav}.
Section \ref{sec:rere} uses this
approximate factorisation to  estimate bounds on individual operators
based on the analysis
of ref.~\cite{CMS1}. The Appendix  collects  the partonic
cross sections for  various  contact interactions. 
 Its aim is to allow an interested reader to
estimate  bounds  from future data on their 
selection of contact interactions,  following
the mostly analytic  recipe  given in section~\ref{sec:rere}.

%%%%%%%%%%%%%%%%%%%%%%%%%%%%%%%%%%%%%%%%%%%%%%%%%%%%%%%%%%%%%%%%%%%%%%
%% SECT  %%%%%%%%%%%%%%%%%%%%%%%%%%%%%%%%%%%%%%%%%%%%%%%%%%%%%%
%%%%%%%%%%%%%%%%%%%%%%%%%%%%%%%%%%%%%%%%%%%%%%%%%%%%%%%%%%%%%%%%%%%%%%

\section{{Kinematics of $pp \to$ dijets}}
\label{sec:cin}

At the LHC, the cross section for $pp \to$ dijets contains
contributions from  QCD,
 electroweak bosons, and possibly from  four-parton contact interactions. 
The purely QCD (or QED) contribution  falls
off as $1/\hats$, where $\hats = M^2_{dijet}$ 
is the four-momentum-squared of the pair of jets,
and grows in the forward/backward directions.
On the other hand,  the contact interaction contribution
grows with $\hats$ and is fairly central.
 So bounds on  contact
interactions can be obtained from the meagre population
of central high energy dijets. 
With a clever choice of variables, the  distribution
in rapidity and  invariant-mass squared of the two jets
can be approximated as the partonic cross section,
multiplying an integral of pdfs. We review here this
approximation.

\begin{figure}[t]
\unitlength.5mm
\SetScale{1.418}
\begin{boldmath}
\begin{center}
\begin{picture}(60,40)(0,0)
\ArrowLine(0,40)(20,25)
\ArrowLine(20,25)(40,40)
\ArrowLine(20,15)(40,0)
\ArrowLine(0,0)(20,15)
\Gluon(20,15)(20,25){2}{2}
\Text(-2,40)[r]{$ k_1$}
\Text(2,33)[r]{$i$}
\Text(42,40)[l]{$p_1$}
\Text(37,32)[l]{$m$}
\Text(42,0)[l]{$ p_2$}
\Text(37,8)[l]{$n$}
\Text(-2,0)[r]{$ k_2$}
\Text(2,8)[r]{$j$}
%\GCirc(20,20){2}{.5}
%\Text(20,20)[c]{$\bullet$}
\end{picture}
\hspace{1cm}
\begin{picture}(60,40)(0,0)
\ArrowLine(0,40)(20,25)
\ArrowLine(20,25)(40,0)
\ArrowLine(20,15)(40,40)
\ArrowLine(0,0)(20,15)
\Gluon(20,15)(20,25){2}{2}
\Text(-2,40)[r]{$ k_1$}
\Text(2,33)[r]{$i$}
\Text(42,40)[l]{$ p_1$}
\Text(37,32)[l]{$n$}
\Text(42,0)[l]{$ p_2$}
\Text(37,8)[l]{$m$}
\Text(-2,0)[r]{$ k_2$}
\Text(2,8)[r]{$j$}
%\GCirc(20,20){2}{.5}
%\Text(20,20)[c]{$\bullet$}
\end{picture}
\hspace{1cm}
\begin{picture}(60,40)(0,0)
\ArrowLine(0,40)(20,22)
\ArrowLine(20,22)(40,40)
\ArrowLine(20,18)(40,0)
\ArrowLine(0,0)(20,18)
\Text(-2,40)[r]{$ k_1$}
\Text(2,33)[r]{$i$}
\Text(42,40)[l]{$ p_1$}
\Text(37,32)[l]{$m$}
\Text(42,0)[l]{$ p_2$}
\Text(37,8)[l]{$n$}
\Text(-2,0)[r]{$ k_2$}
\Text(2,8)[r]{$j$}
\GCirc(20,20){2}{.5}
%\Text(20,20)[c]{$\bullet$}
\end{picture}
\hspace{1cm}
\begin{picture}(60,40)(0,0)
\ArrowLine(0,40)(20,22)
\ArrowLine(20,22)(40,0)
\ArrowLine(20,18)(40,40)
\ArrowLine(0,0)(20,18)
\Text(-2,40)[r]{$ k_1$}
\Text(2,33)[r]{$i$}
\Text(42,40)[l]{$ p_1$}
\Text(37,32)[l]{$n$}
\Text(42,0)[l]{$ p_2$}
\Text(37,8)[l]{$m$}
\Text(-2,0)[r]{$ k_2$}
\Text(2,8)[r]{$j$}
\GCirc(20,20){2}{.5}
%\Text(20,20)[c]{$\bullet$}
\end{picture}
\end{center}
\end{boldmath}
\caption{\label{figQCDLam}
Possible 
QCD and contact interaction diagrams
for    $q_iq_j \to q_mq_n$.
QED diagrams with the gluon replaced by
a photon are also possible. 
At the grey blob representing
the contact interaction, the quark lines may
have chiral projectors and/or a {colour} matrix.
 Which gauge diagrams interfere with a given 
contact interaction
will depend on its flavour indices and operator structure; for instance, for $V-A$
operators with
$i = j= n = m$,  all four gauge diagrams could interfere.
 }
\end{figure}
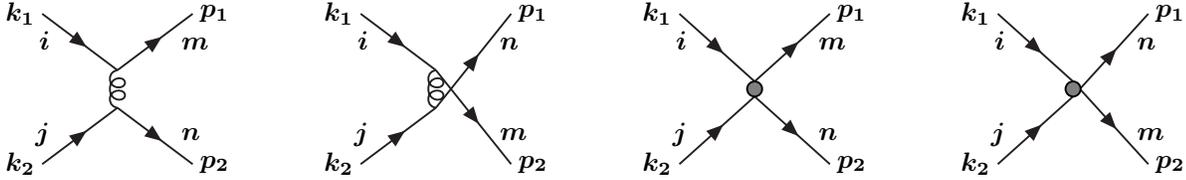

The parton-level diagrams for quark-quark scattering
 $q_i(k_1) q_j(k_2) \to q_m(p_1) q_n(p_2)  $
 are given in Figure~\ref{figQCDLam}, where $i,j,m,n$ are flavour indices,
and $\{k,p\}$ are four-momenta.  
The resulting cross sections, which can be
parametrised with the partonic mandelstam
variables $\hats, \hatt$, and $\hatu$  are listed in the
Appendix. 
To obtain an observable, 
the partons must  be embedded in the incident protons,
here taken to have four-momenta $P_\pm$. {We denote} $f_j(x_1)$  the  probability
density  that
the parton $j$ carries a fraction $x_1$ of  the four-momentum of
an incident  proton, so that
$\hats = x_1x_2 (P_+ +P_-)^2$.   The 
 total cross section  can be written
\beq
\sigma (pp \to  dijets) = 
\sum_{i,j,m,n} ~
\int_0^1 d x_1
\int_0^1 d x_2
 f_i(x_1)  f_j(x_2) \sigma (i(x_1 P_+) j(x_2 P_-)
\to mn) 
\eeq
where  the sum {runs} over all possible incident partons for
each partonic process, and over  all the partonic processes
which contribute ($e.g.$, in the case
of dijets, $ qq \to qq, q \bar{q}
\to q \bar{q}, gg \to q \bar{q},gg \to gg$, etc).
In our estimates, we
only include   $uu \to uu, dd\to dd$, $ud\to ud$  
and  $ug\to ug$  
in the partonic QCD cross section. These should
be the main contributions to the dijet cross section.
The partonic cross sections increase by
a factor $\sim 9/4$ for each initial gluon:
$\sigma(qq\to qq):\sigma(qg\to qg):\sigma(gg\to gg)
\sim \frac{4}{9}:1:\frac{9}{4}$.
However,
the  density of gluons in the proton, at the
large values of $x$ which are relevant {here}, is 
at least a factor of 1/10 (1/3) below that of
valence  $u$ ($d$) quarks, and the density
of sea quarks, is two orders of magnitude below the $u$
density, {which justifies our approximation.}

It can be  convenient to 
introduce the pseudo-rapidities
\beq
\label{y}
y =  \frac{1}{2} \ln \frac{E+ p_z}{E-p_z}~~
\eeq
of the individual jets, 
  and
their  combined mass squared
\beq
M^2_{dijet} = \hats = x_1 x_2  s ~~.
\label{M2dijet}
\eeq
We interchangeably  refer to $M^2_{dijet}$ or  $\hats$ 
throughout the paper, {using the convention that}
partonic variables, such as the Mandelstam $\hats,\hatt,\hatu$
(defined in eq.~(\ref{mandelstam})),  wear hats. 

The  CMS search~\cite{CMS1} for contact interactions
in the angular distribution of dijets at high $M^2_{dijet}$
uses as variables the dijet mass squared eq.~(\ref{M2dijet}), 
  the pseudo-rapidity of the
partonic centre of mass frame  \beq
y_+ = (y_1 +y_2)/2 ~~~,
\label{y+}
\eeq
and 
\beq
\chi = \exp| y_1 - y_2| = \frac{1 + |\cos \theta^*|}{1 - |\cos \theta^*|}      ~~,
\label{chi}
\eeq
where $\theta^*$ is the centre-of-mass scattering
angle  away from the beam axis (see the Appendix). 

At the parton level, the  
QCD  contributions  to the
differential cross section ${d \hat{\sigma}}/{d\hatt}$
have a $1/\hatt^2$ divergence, {leading to a}
large rate for
small-angle scatterings along the beam pipe. 
On the other hand, the dijets produced by contact 
interactions have a more isotropic distribution. 
The most sensitive place to look for contact interactions
is therefore in large-angle scatterings, producing
dijets in the central part of the detector. Expressed as a function of $\chi$,
the QCD contribution to the dijet cross section
is approximately flat,
whereas the contact interaction contribution
peaks at small $\chi$. This is illustrated
in figs.~1 and 2 of ref.~\cite{CMS1}, where the main
effect of contact interaction occurs for
$1\leq \chi \leq 3$ bin, which
corresponds to $60^o \leq \theta^*\leq 120^o$.

With  these
variables, 
the differential dijet cross section is 
\bea
\frac{d \sigma}{d y_+~d \chi ~ d M_{dijet}^2 } 
&=&\frac{   f_i(x_1) f_j(x_2)}{s} 
\frac{d \hat{\sigma}}{d \hatt}  
\frac{\hatt^2}{\hats}~~~.
\label{Q1}
\eea
For fixed dijet mass,
the pdfs depend on $y_+$ but not $\chi$ 
 (since 
$x_1 = e^{y_+}{M_{dijet}}/{\sqrt{s}} ,
x_2 = e^{-y_+}{M_{dijet}}/{\sqrt{s}}  $),
 and the
partonic differential cross section depends on
$\chi$ but not $y_+$, so the expected number of events
can be factorised  as  a integral-of-pdfs,
multiplied by a partonic cross section. 
 Therefore, in an ideal world with contact interactions,
the dijet distribution in mass and $\chi$ could allow one
to determine the actual partonic cross section, 
providing the necessary information
to identify the
operator(s) that  induced it.

\section{Contact interactions: operators versus effective interactions}
\label{sec:opinteff}

\subsection{Operators}
\label{sec:op}

New Physics from   a scale $M > m_W$
can be described,  at scales $\ll M$, by 
an effective Lagrangian containing the 
renormalisable Standard
Model (SM) interactions, and   
various  $SU(3) \times SU(2) \times U(1)_Y$-invariant 
operators of dimension $>4$, with coefficients
determined by the New Physics model.  There can be relations
among the operators, arising from
symmetries and equations of motion. 
This section gives 
a  basis of
 dimension six,  four-quark operators, taken
from  the Buchmuller-Wyler~\cite{BW} list as pruned by ref.~\cite{polonais}~\footnote{We restrict {our analysis} to dimension-6 operators. This is a reasonable
perturbative approximation when the next-order
terms, relatively suppressed by $\sim \hats/\Lambda^2, v^2/\Lambda^2$
can be neglected --- which appears to be barely the case here.}.

In the following list of four quark operators,
 $Q$ are electroweak  doublets, $U$ and  $D$  are singlets,
$\lambda^A$  are the generators of  SU(3), $\vec{\tau}/2$ are
those of SU(2), the gauge index sums are implicit
inside the parentheses, and  $i,j,m,n$  are generation indices which
all run from 1\ldots3 (so there is a 
 factor 1/2 in front of operators made of same
current twice, to obtain the Feynman rule {described below} in eq.~(\ref{Frule})).
\bea
{\cal O}^{1,1}_{\bar{Q} Q} =
\frac{1}{2}
(\overline{Q}_m \gamma^\mu Q_i ) (\overline{Q}_n \gamma_\mu Q_j )  &&
\label{BW1}
\\
{\cal O}^{1,3}_{\bar{Q} Q} =
\frac{1}{2}
(\overline{Q}_m \gamma^\mu \vec{\tau}  Q_i ) 
(\overline{Q}_n \gamma_\mu \vec{\tau}  Q_j )   &&
\label{BW2}  \\
{\cal O}^{1,1}_{\bar{U} U} =
\frac{1}{2}
(\overline{U}_m \gamma^\mu U_i ) (\overline{U}_n \gamma_\mu U_j )  &&
\label{BW3} \\
{\cal O}^{1,1}_{\bar{D} D} =
\frac{1}{2}
(\overline{D}_m \gamma^\mu D_i ) (\overline{D}_n \gamma_\mu D_j )  &&
\label{BW4}
\eea
There are also  operators contracting  currents of singlet quarks of
different charge:
\bea
{\cal O}^{1,1}_{\bar{U} D} =
(\overline{U}_m \gamma^\mu U_i ) (\overline{D}_n \gamma_\mu D_j )  &&
{\cal O}^{8,1}_{\bar{U} D} =
(\overline{U}_m \gamma^\mu \lambda^A U_i ) 
(\overline{D}_n \gamma_\mu \lambda^A D_j )
\label{BW5} 
\eea
and operators contracting doublet and singlet currents:
\bea
{\cal O}^{1,1}_{\bar{Q} D} =
(\overline{Q}_m \gamma^\mu Q_i ) (\overline{D}_n \gamma_\mu D_j )  &&
{\cal O}^{8,1}_{\bar{Q} D} =
(\overline{Q}_m \gamma^\mu \lambda^A Q_i ) 
(\overline{D}_n \gamma_\mu \lambda^A D_j ) \label{opex1} \\
{\cal O}^{1,1}_{\bar{Q} U} =
(\overline{Q}_m \gamma^\mu Q_i ) (\overline{U}_n \gamma_\mu U_j )  &&
{\cal O}^{8,1}_{\bar{Q} U} =
(\overline{Q}_m \gamma^\mu \lambda^A Q_i ) 
(\overline{U}_n \gamma_\mu \lambda^A U_j )  \label{opex2}
\eea
and finally there is a scalar operator, with 
antisymmetric SU(2)
index contraction across the parentheses:
\bea
{\cal O}^{S,1,1}_{\bar{Q}  \bar{Q}} =
(\overline{Q}_m  U_i ) (\overline{Q}_n D_j )  &&
{\cal O}^{S,8,1}_{\bar{Q} \bar{Q} } =
(\overline{Q}_m \lambda^A U_i ) 
(\overline{Q}_n  \lambda^A D_j )  \label{BWscal3} ~~.
\eea

\subsection{Effective interactions}
\label{sec:inteff}

Once electroweak symmetry is broken,
a particular operator induces one or several  effective interactions among  mass
eigenstates.
We aim at constraining these effective interactions, one at a time, through
the LHC searches on contact interactions.

This leads us to make
 a distinction between
(gauge-invariant) operators, and effective interactions
(having distinct external legs).
The aim of this distinction is to address
a general problem with setting bounds on the coefficients
of gauge-invariant operators~\cite{AEN}: such  bounds
may not transfer, in a simple way, from one
operator basis to another.
 The selection
of SM  gauge-invariant operators made in
section~\ref{sec:op} is not unique,
as expected for a {choice of} basis.
A different list of operators
might be more suited to describing some 
models {because they capture the symmetries 
of the model in a more economical way}. 
Ideally,  the constraints  on four-fermion operator
coefficients  should be transferable from one
 operator basis to another. However, this is
not possible if 
these constraints are obtained by turning on
one operator at a time, as will be done here.
One difficulty is that an operator,
such as the SU(2) triplet of eq.~(\ref{BW3}), induces
several four-fermion interactions,
so  quoting a bound on the operator
loses the {correlation carrying the} information about which interaction
the bound arose from.
In addition,  if the operators can interfere among
themselves  
(for instance, 
the singlet and triple operators of
eqs.~(\ref{BW1}) and (\ref{BW3})  
can interfere), then the bound
on a sum of operators could be less restrictive
than the bounds on single operators.

To circumvent the problem
of ``basis-dependent'' bounds~\footnote{
Ref.~\cite{NLO} shows some parameter choices where
interference among operators could  reduce  
sensitivity.},
we follow ref.~\cite{Carpentier}, and set  limits  on 
the coefficients of ``effective four-quark interactions'',
These effective interactions should  be distinct, 
so that they do not interfere among each other, 
and should include all the interactions
induced by the effective operators. Unfortunately,
these two requirements are not quite compatible; 
there are two interactions which can
interfere, but we neglect this effect.

We obtain the list by considering  the
$U(1) \times SU(3)$ invariant  operators generated  after electroweak
symmetry breaking by the previous basis of SM gauge
invariant operators {(decomposing the SU(2)
doublets $Q$ into their components)}. 
These 
interactions almost never  interfere among themselves
(the exceptions are eqs.~(\ref{opeffCC}) and (\ref{opeffscal})), 
which ensures that
the prediction of a sum of interactions
will be the sum of the predictions of the 
interactions taken separately. 
The possibilities are the following:
\begin{itemize}
\item the ``neutral-current'' 
 left-left or right-right  interactions, $X = L$ or $R$:
\bea
{\cal O}^{1,XX}_{u_mu_iu_nu_j} =
\frac{1}{2}
(\overline{u}_m \gamma^\mu P_X u_i ) (\overline{u}_n \gamma_\mu P_X  u_j )  &&
\label{op1} \\
{\cal O}^{1,XX}_{d_md_id_nd_j} =
\frac{1}{2}
(\overline{d}_m \gamma^\mu P_X d_i ) (\overline{d}_n \gamma_\mu P_X  d_j )  &&
\label{op2}\\
{\cal O}^{1,XX}_{u_mu_id_nd_j} =
(\overline{u}_m \gamma^\mu P_X u_i ) (\overline{d}_n \gamma_\mu P_X  d_j )  &&
{\cal O}^{8,RR}_{u_mu_id_nd_j} =
(\overline{u}_m \gamma^\mu  P_R \lambda^A u_i ) 
(\overline{d}_n \gamma_\mu   P_R \lambda^A d_j ) ~~~.
\label{op3}
\eea
Notice that, with  the previous basis
of gauge-invariant operators,
 the octet ${\cal O}^{8,RR}_{uudd}$ only arises for singlet
currents.
\item the ``charged-current'' interactions
\bea
{\cal O}^{1,CC} =
(\overline{u}_m \gamma^\mu P_L d_i ) (\overline{d}_n \gamma_\mu P_L  u_j )  &&
{\cal O}^{8,CC} =
(\overline{u}_m \gamma^\mu  P_L \lambda^A d_i ) 
(\overline{d}_n \gamma_\mu   P_L \lambda^A u_j ) ~~~,
\label{opeffCC}
\eea
the second of which can be rearranged to
a linear combination of ${\cal O}_{u_md_id_nu_j}^{1,CC}$
and ${\cal O}_{u_mu_jd_nd_i}^{1,LL}$. 
 We therefore do not include octet
charged-current 
interactions. 
We include the singlet ${\cal O}_{u_md_id_nu_j}^{1,CC}$,
which unfortunately can interfere with
 ${\cal O}_{u_mu_jd_nd_i}^{1,LL}$ (see eq.~(\ref{dsigdt6cCC})).
\item
the ``neutral-current'' left-right  operators,
\bea
{\cal O}^{1,XY}_{u_mu_iu_nu_j} =
\frac{1}{2}
(\overline{u}_m \gamma^\mu P_X u_i ) (\overline{u}_n \gamma_\mu P_Y  u_j )  &&
{\cal O}^{8,XY}_{u_mu_iu_nu_j} =
\frac{1}{2}
(\overline{u}_m \gamma^\mu  P_X \lambda^A u_i ) 
(\overline{u}_n \gamma_\mu   P_Y \lambda^A u_j ) \\
{\cal O}^{1,XY}_{d_md_id_nd_j} =
\frac{1}{2}
(\overline{d}_m \gamma^\mu P_X d_i ) (\overline{d}_n \gamma_\mu P_Y  d_j )  &&
{\cal O}^{8,XY}_{d_md_id_nd_j} =
\frac{1}{2}
(\overline{d}_m \gamma^\mu  P_X \lambda^A d_i ) 
(\overline{d}_n \gamma_\mu   P_Y \lambda^A d_j ) \\
{\cal O}^{1,XY}_{u_mu_id_nd_j} =
(\overline{u}_m \gamma^\mu P_X u_i ) (\overline{d}_n \gamma_\mu P_Y  d_j )  &&
{\cal O}^{8,XY}_{u_mu_id_nd_j} =
(\overline{u}_m \gamma^\mu  P_X \lambda^A u_i ) 
(\overline{d}_n \gamma_\mu   P_Y \lambda^A d_j ) %\\
\eea
where $ P_X,P_Y \in \{ P_L,P_R \}, X\neq Y.$
\item the scalar operators {stemming from} eq.~(\ref{BWscal3}), 
which each give two interactions:
\bea
{\cal O}^{S1}_{q_m u_i q_nd_j}& =&
(\overline{d}_m P_R u_i ) (\overline{u}_n P_Rd_j )
-(\overline{u}_m P_R u_i ) (\overline{d}_nP_R d_j )\nonumber \\
{\cal O}^{S8}_{q_m u_i q_nd_j}& =&
(\overline{d}_m P_R\lambda^A u_i ) 
(\overline{u}_n P_R  \lambda^A d_j ) 
-
(\overline{u}_m P_R \lambda^A u_i ) 
(\overline{d}_n P_R \lambda^A d_j )  \label{opeffscal} ~~.
\eea
For the scalar operators, we constrain the coefficients
of the operators,  that is, of ${\cal O}^{S1}$
and ${\cal O}^{S8}$ given above,  because the interactions
in the sums interfere between each other. 
\end{itemize}

\subsection{Feynman rules and dimensional analysis}
\label{sec:Frule}

Following the convention
of  collider {constraints on contact interactions},
we suppose that the coupling in the four-quark
Feynman rule is 
\beq
 i \eta \frac{ 4 \pi }{\Lambda^2} ~~~~, ~~~
{\rm with}~ \eta = \pm 1~.
\label{Frule}
\eeq
The  operators/interactions,
of  the
previous sections are  normalised such 
that their coefficient  should  be $\eta4\pi/\Lambda^2$.
For each flavoured contact interaction,
there will be a different lower  bound on $\Lambda$.
If the contact interaction is generated perturbatively
by the exchange of a particle of mass $M$ with coupling $g'$, 
one would expect a contact interaction like $g^{'2}/M^2$,
{leading to the scale} $\Lambda\sim M/\sqrt{\alpha'}$.

In the matrix element squared, 
the contact interaction may interfere with
QCD (depending on its colour, flavour and
chiral structure) as well as  with itself. The
$pp \to$ dijet cross section, in the central
part of the detector,  is therefore of the 
form
\beq
\frac{d\sigma}{d\chi} \sim
C_{QCD}\frac{\alpha_s^2}{\hats}
+ C_{2}\frac{\alpha_s}{\Lambda^2}
+ C_{4}\frac{\hats}{\Lambda^4}~~,
\label{soln}
\eeq
where the $C_x$ are  ${\cal O} (1)$ constants that depend on
the {specific} contact interaction. {$C_4$, related to the contribution
of contact interaction alone, is always positive and will dominate
for very large $\hat{s}$, whereas $C_2$, stemming from
the interference between QCD and contact interactions, can be positive or
negative depending on the value of $\eta$, and can thus induce
either a deficit or an excess of events at intermediate values of $\hat{s}$.}
Requiring that the contact
interactions induce a deviation of $ \epsilon \lsim 1$
from the QCD expectation, we need 
\beq
\frac{\hats}{\alpha_s} \sim \Lambda^2 ~~.
\label{ordgran}
\eeq
{suggesting that} for $\sqrt{\hats}= 3$ TeV at the 7-8 TeV LHC~\cite{CMS1}, the limits on
$\Lambda$ should be $O(10)$ TeV. {In the following, we will impose a tighter constraint 
on eq.~(\ref{soln}) for specific contact interactions, providing different bounds depending on their flavour
structure.}

\section{Contact interactions: flavour structure}
\label{sec:sav}

\subsection{{Impact of flavour on the search for contact interactions}}

Each of the interactions given in the  section~\ref{sec:inteff} exists
in  a plethora of flavoured combinations, only some of which
can be constrained  by colliders.
The flavour indices  have two effects on  the  collider bounds: first,
some flavours are more plentiful in the proton
than others,  and
second,  the  cross section can involve interferences 
with QCD  or QED, which depend on  the flavours.

{First,} at the LHC, bounds can be set on
contact interactions with two incoming valence quarks
($uu, dd, ud$). This is because the density
of sea quarks and antiquarks is very suppressed.
The incident partons should both carry a significant
fraction of the proton momentum, to produce
 a pair of  jets of large combined mass (the last bin in dijet
mass of the CMS analysis~\cite{CMS1} is $M_{dijet}> 3$TeV 
at the 7 TeV LHC, corresponding to $x_1 x_2 \geq 9/49$).
At such large values of $x$, the density of 
all flavours of sea quark and anti-quark is {of similar size}
and about two orders
of magnitude below the density of $u$ {quarks}. 
It is therefore doubtful, with current data,
to set a bound on contact interactions with an
initial sea quark or anti-quark, because 
a huge cross section would be required to compensate
the {relative suppression of} sea pdfs.

The second effect of flavour indices is on the partonic
cross section. For instance, the cross section
for $ud \to ud$, mediated by QCD + contact interactions,
is different from $uu \to uu$ (see eqs.~(\ref{P+S17.71})-(\ref{P+S:17.64})). 
In particular,   the singlet
interactions mediating $uu\to uu$  interfere with QCD,
whereas there is not interference {between} QCD and
singlet interactions mediating  $ud \to ud$. 
Flavour-summed contact interactions
constrained by the experimental collaborations {have thus a reduced sensitivity to}
destructive interferences {as they get contributions from several flavour operators, most of them being
positive contributions coming from contact interactions alone that increase 
the cross sections}.
In the Appendix,
we attempt to classify the possible flavour index
combinations, and  provide  their partonic cross sections.

Another comment is in order at this stage. The experimental final state considered here is a pair of jets,
so the final state quarks can be of any flavour other
than top.  This means  the LHC  is sensitive to curious
$\Delta F = 2$ and $\Delta F = 1$ flavour-changing  
operators mediating processes
like $uu \to cc$ or $ud\to ub$. Flavour physics 
({\it e.g.} meson mixing and $B$ decays) can impose more stringent bounds
on some of them.
{In the  following, we will} give the LHC bounds on the
various  flavour-changing operators.

\subsection{Comparing to the existing notation}
\label{ssec:comp}

Traditionally~\cite{ELP},  collider searches for contact interactions
quote bounds on the mass scale $\Lambda$
appearing in an interaction of  the form:
\beq
{\cal L}_{Pythia} = \frac{4 \pi}{\Lambda^2}
\sum_{i,j = u,d,s,c,b}
\left[
\frac{\eta_{LL}}{2}
(\overline{q}_i \gamma^\mu P_L q_i ) (\overline{q}_j \gamma_\mu P_Lq  _j )
+
\frac{\eta_{RR}}{2}
(\overline{q}_i \gamma^\mu P_R q_i ) (\overline{q}_j \gamma_\mu P_R  q_j )
+
\eta_{LR}
(\overline{q}_i \gamma^\mu P_R q_i ) (\overline{q}_j \gamma_\mu P_L  q_j )
\right] ~~.
\label{inpy8}
\eeq
Specifically, this is the contact 
interaction coded into {\sc  pythia} 8~\cite{pythia8},
where the $\eta$ coefficients can be chosen to be $\pm1$ or 0.
Some frequently 
studied combinations of $\eta$ coefficients  are given in table \ref{tab:eta}.
\renewcommand{\arraystretch}{1.5}
\begin{table}[t]
\begin{center}
\begin{tabular}{|c|c|c|}
\hline
& $(\eta_{LL},\eta_{RR},\eta_{LR}) $ & operators \\
 \hline
$\Lambda_{LL}^{\pm}  $
&$(\pm 1 ,0 ,0 )$ 
& $ \pm[{\cal O}^{1,LL}_{uuuu} + 2  {\cal O}^{1,LL}_{uudd} 
+ {\cal O}^{1,LL}_{dddd}]$  \\
$\Lambda_{RR}^{\pm } $& 
$ (0 ,\pm 1 , 0)$
&  $\pm[{\cal O}^{1,RR}_{uuuu} +2 {\cal O}^{1,RR}_{uudd}
+ {\cal O}^{1,RR}_{dddd}] $\\
$\Lambda_{VV}^{\pm }$
&$ (\pm 1 ,\pm 1 ,\pm 1 )$
& $
\pm[ \sum_{m,n = u,d} ( {\cal O}^{1,LL}_{mmnn}
+{\cal O}^{1,LR}_{mmnn}
+{\cal O}^{1,RL}_{mmnn}
+{\cal O}^{1,RR}_{mmnn} )]$
% , {\cal O}^{1,LL}_{dddd} +  {\cal O}^{1,LR+RL}_{dddd} + {\cal O}^{1,RR}_{dddd} $
 \\
$\Lambda_{AA}^{\pm } $&
$ (\pm 1 ,\pm 1 ,\mp 1 )$
&$
\pm[ \sum_{m,n = u,d} (
{\cal O}^{1,LL}_{mmnn}
-{\cal O}^{1,LR }_{mmnn}
-{\cal O}^{1, RL}_{mmnn}
+{\cal O}^{1,RR}_{mmnn})] 
$
% , {\cal O}^{1,LL}_{dddd} -  {\cal O}^{1,LR+RL}_{dddd} + {\cal O}^{1,RR}_{dddd} $
\\
$\Lambda_{V-A}^{\pm } $&
$ (0 , 0,\pm 1 )$
& $\pm[{\cal O}^{1,LR=RL}_{uuuu}+{\cal O}^{1,LR}_{uudd}+{\cal O}^{1,LR}_{dduu} 
+  {\cal O}^{1,LR=RL}_{dddd}] $\\
\hline
\end{tabular}
\caption{Sums of operators contributing (at the LHC) to some 
commonly studied contact interactions (see eq.~(\ref{inpy8})).
On the left, the sub/super-scripts for $\Lambda$
indicate
the choice of $\eta$ coefficients. On the right are given the
corresponding interactions from section~\ref{sec:inteff}.
In principle, contact interactions studied in collider experiments
have a sum over all flavours; however, the LHC is principally
sensitive to contact interactions with two incoming valence
quarks, so the flavour sum is over $u,d$.
\label{tab:eta}}
\end{center}
\end{table}
\renewcommand{\arraystretch}{1}

With the aim of obtaining useful and conservative bounds,
our constraints will differ in three ways:
\ben
\item We include 
the octet operators ${\cal O}^8$ of section~\ref{sec:inteff}.
They generally
give smaller modifications to  the dijet rate, so
the bound on $\Lambda$ is lower, as can be seen 
from the tables of section~\ref{sec:rere}.

\item  We constrain each combination of flavour indices
separately.

\item We only consider 
 ${\cal O}^{1,LR} $ (or ${\cal O}^{1,RL} $),
 and 
$ {\cal O}^{1,LL} $ (or ${\cal O}^{1,RR} $),
but not the various linear combinations available
in table \ref{tab:eta}.

\een

Turning on  one  effective interaction of
given flavours   at a time (as done here) gives conservative bounds
if two conditions are satisfied. First, the
interactions should not interfere among themselves,
so that the contribution of the sum is
the sum of the contributions. This is  almost
the case for the  interactions
of section~\ref{sec:inteff}.
 Second, each bound
should arise from requiring that the
operator not induce an excess of events
(as opposed to a deficit of events). This will be
true for the bounds that we will derive.

Suppose now  that one wishes to set a  bound on
a specific New Physics model, which induces
a sum of low-energy contact interactions ---
for instance the $VV$ combination. 
It is simple and conservative
to  take the strongest of the  bounds obtained 
one-at-a-time  for the operators in the sum. 
However, the true limit should be better.
It is not straightforward  
to obtain the bound on  $\Lambda_{VV}^+$
given the limits on
 $\Lambda_{XX}^+$ ($X = R$ or $L$) and   
 $\Lambda_{LR}^+$.
Despite that the  excess events induced by 
$ {\cal O}^{1,VV} $ are the sum of the
excesses due to 
 ${\cal O}^{1,XY} $  and 
$ {\cal O}^{1,XX} $ 
(for $X\neq Y \in \{ L,R\})$,
 there are two hurdles to
obtaining a  bound  on  $\Lambda_{VV}^+$:
first, one must calculate 
the  partonic  cross section
for the $ {VV} $ operator combination, 
then one must know how it is constrained
by the data. To address the
first hurdle,  we collect in the Appendix 
the partonic cross sections for a variety of
contact interactions.  The second hurdle
is a problem, because it is clear that
the experimental collaborations cannot
constrain all possible combinations of all
contact interactions. However,
contact interactions induce excess
high-mass dijets in the central part of the
detector, so  observables which
measure this,  such
as $F_\chi(M_{dijet})$~\cite{ATLAS} of the ATLAS
collaboration, should be translatable to  limits on
generic contact interactions.

%%%%%%%%%%%%%%%%%%%%%%%%%%%%%%%%%%%%%%%%%%%%%%%%%%%%%%%%%%%%%%%%%%%%%%
%% SECT  %%%%%%%%%%%%%%%%%%%%%%%%%%%%%%%%%%%%%%%%%%%%%%%%%%%%%%
%%%%%%%%%%%%%%%%%%%%%%%%%%%%%%%%%%%%%%%%%%%%%%%%%%%%%%%%%%%%%%%%%%%%%%

\section{Estimating bounds on flavoured contact interactions}
\label{sec:rere}

\subsection{From data to partonic cross sections}%Partonic cross sections}   \label{sec:xc}
\label{ssec:From}

Suppose that an effective interaction has been selected,  
with  incident flavours  of the first
generation.
The recipe to guess a bound on $\Lambda$, from 
the dijet distribution in                   
 $M^2_{dijet}$  and $\chi$,  is simple:
\begin{enumerate}

\item  look up  the (flavoured)
partonic cross section  for  the selected  
contact interaction plus  QCD,
and evaluate at the $\chi$  corresponding to the bin.

\item integrate the  pdfs
of the incident partons over the $y_+$ values
which  are consistent with the experimental cuts.

\item  {multiply 1)  by 2)}, and require that it
agree with the QCD expectation  for  the bin.

\end{enumerate}
{We want to illustrate in detail the procedure  
in the case of the CMS analysis~\cite{CMS1}.} 
 The CMS collaboration  measured the distribution
of dijets in $\chi$ from $1\to 16$, and $M_{dijet}$ from 0.4 to
$\geq 3$ TeV~\cite{CMS1}.
We focus on the  highest  $M_{dijet}> 3$ TeV bin
(obtained with 2.2 fb$^{-1}$ of data), 
for which CMS plots the   normalised~\footnote{{At this stage, we}
suppress the $M^2_{dijet}$ dependence; 
 $\frac{d \sigma}{d \chi}$ means $
\int dy_+ \frac{d^3 \sigma}{d y_+d M^2_{dijet} d \chi}$, and
$\sigma =  \int d\chi dy_+  \frac{d^3 \sigma}{d y_+d M^2_{dijet} d \chi}$.
} differential
cross sections 
$( \equiv \frac{1 }{\sigma }\frac{d \sigma}{d \chi})$,
corresponding to the data, to the QCD expectation, and
to  the predictions of QCD plus  contact interactions ({denoted} QCD+CI).
{As indicated in section~\ref{sec:cin}, the highest sensitivity to contact interactions is obtained for} 
the $1\leq \chi\leq 3$ bin.
A ratio which can be  extracted 
for {a specific} bin  is
\beq
%\frac{
\frac{1}{\sigma_{QCD+CI} }\frac{d \sigma_{QCD+CI}}{d \chi} 
\left[
\frac{1}{\sigma_{QCD}}\frac{d \sigma_{QCD}}{d \chi}\right]^{-1} ~~~,
\label{exp}
\eeq
where all the cross sections are for $ pp\to$ dijets
(partonic cross sections will wear hats):
\beq 
\frac{d \sigma_{QCD+CI}}{d \chi} =
\sum_{i,j,m,n} \int d y_+ f_i f_j \, \left(
 \frac{d \hat{\sigma}_{QCD}}{d \chi}
(ij \to mn)% (q_i q_j \to  q_m q_n)
+ \frac{d \hat{\sigma}_{QCD*CI}}{d \chi}
(ij \to mn)% (q_i q_j \to  q_m q_n)
+ \frac{d \hat{\sigma}_{CI}}{d \chi}
 (ij \to mn)%(q_i q_j \to  q_m q_n)
\right) ~~,
\label{dsdtphys} \eeq
and are summed over the $\chi$ and
$\hats$ {ranges} of the bin (we will return
to these sums later).

Clearly,  the  prediction of
contact interactions  should be compared with 
the data,  not with  the QCD prediction. However, we notice that
the data~\cite{ATLAS,CMS1} agree with the QCD prediction
  (for $M_{dijet} > 3$ TeV they are marginally 
below the predictions for low-$\chi$ bins and above for 
  high-$\chi$ bins, in the case of ref.~\cite{CMS1}).
So we   ``normalise'' our (incomplete) leading
order  parton-level  QCD cross section, to the
QCD expectation obtained by ATLAS and CMS
at next-to-leading order (NLO) with hadronisation and detector effects.
Then  we estimate   bounds on
contact interactions  by requiring that
they add $\lsim 1.6 \sigma$ to the QCD  contribution,
where $\sigma$ is the  experimental
statistical and systematic uncertainties
for the relevant bin, added in quadrature. 
{On the basis of the results of ref.~\cite{CMS1},} we estimate that this allows contact interactions to
contribute {from 1/3 to 1/2} of the QCD contribution 
{either positively or negatively. In other words, even though we base our 
analysis on the most sensitive $\chi$-bin (from 1 to 3) at
maximal $M_{dijet}$, which 
has an observed value slightly below the QCD prediction, 
we consider that the spread of data with respect to QCD 
in the other bins prevents us from interpreting the 
deficit in the 1-3 bin as a negative contribution 
from contact interactions. We thus take the more conservative 
approach to set a bound on contact interactions as a 
fraction (positive or negative) from QCD 
(we will come back to this point in section~\ref{sec:slumi}).}

The  ratio (\ref{exp}) 
can be related, in a series of steps, to a ratio  of partonic 
cross sections. 
\begin{itemize}
\item \emph{The first step} {consists in canceling}
$ \sigma_{QCD+CI} \simeq  \sigma_{QCD}$ {in the ratio}. This  is a  self-consistent
approximation, because  contact interactions 
  only contribute in the low   $\chi$ bins, where
they are bounded  to be a fraction of QCD.  

\item \emph{The second step} {amounts to writing} the ratio 
\beq
%\frac{
\frac{d \sigma_{QCD+CI}}{d \chi} 
\left[
\frac{d \sigma_{QCD}}{d \chi}\right]^{-1}  = 1 + \epsilon
\label{exp2}
\eeq
where 
\beq
\epsilon = 
\frac{ \sum_{m,n} \int d \hats \int d y_+ f_i f_j \,  \delta \frac{d \hat{\sigma}}{d \chi}
%_{QCD*CI + CI^2}
 (q_i q_j \to  q_m q_n)}
{\sum_{i,j,m,n} \int d \hats \int d y_+ f_i f_j   \frac{d \hat{\sigma}_{QCD}}{d \chi} 
(q_i q_j \to  q_m q_n)} ~~~,
\label{epsilon}
\eeq
{where}
the partonic differential cross
section $\delta \frac{d \hat{\sigma}}{d \chi}$  is the
modification  to $(q_i q_j \to  q_m q_n)$ 
induced by contact interactions ({alone or through interference with QCD, corresponding to}
the last two terms inside
the parentheses of eq.~(\ref{dsdtphys})).
{The denominator amounts to the QCD contribution which}
is summed over the possible
initial flavour combinations
{(limited to $uu$, $dd$ , $ud$ and $ug$).}
The integral over the  $\hats = M^2_{dijet}$
range of the bin has been reinstated.

\item {\emph{The third step} corresponds to factorising the integrals in the ratio $1+\epsilon$.}
As discussed in section~\ref{sec:cin}, the partonic cross sections 
depend on  $M^2_{dijet}$  and  $\chi$,
and the pdfs depend on $M^2_{dijet}$ and  $y_+$.
We now want to factor the partonic
cross section out of  
 the integral  across  the  $M_{dijet}>3 $ TeV
bin. 
This will be an acceptable
approximation, because  the pdfs drop rapidly with  large
increasing $x$, so contribute most of the
integral in a narrow range of
  $M_{dijet} \sim 3$ TeV. 
  
We factorise
in two steps, starting with the denominator of eq.~(\ref{epsilon}).
All the QCD cross sections scale as $1/\hats$,
so the cross sections evaluated at
$\hats_{min} = (3$ TeV)$^2$ can be factored
out of the integrals and replaced by 
$\hats_{min}/\hats$ {as far as the $s$-dependence is concerned (see Appendix)}.
We are left to integrate for $N=-1$
\beq
I^N_{ij} = \int_{min}^{max} d \hats \left(\frac{\hats}{\hats_{min}}\right)^N \int dy_+ f_i(x_1) f_j(x_2)
~~,~ x_1 = e^{y_+}\sqrt{\frac{\hats}{s}}
~~,~ x_2 = e^{-y_+}\sqrt{\frac{\hats}{s}}~,
\label{IN}
\eeq
over 
the $y_+$ region consistent with the CMS cuts ($y_+ < 1.1$) and the
value of $M^2_{dijet}$ ({corresponding to} $2y_+ < \ln {s}/{\hats}$),
and  over  the range of energy  from $\hats_{min} = (3$ TeV$)^2$ to $\hats_{max} = (6$ TeV$)^2$.  
We use CTEQ10~\cite{CTEQ10} pdfs {(at NLO)} at a scale of 3 TeV.
The results, normalised to $I^{-1}_{uu}$,
are given  in table 
\ref{tab:rij}. 
The ratios in the table  change by only
a few percent  when we change $\hats_{max}$
to 4.5 TeV. 

The second step is to {take}
$\delta \frac{d \hat{\sigma}}{d \chi}$,
evaluated at $\hats = (3$ TeV)$^2$, out
of the  integral {in the numerator of} eq.~(\ref{epsilon}).
We have then to
evaluate $I_{ij}^N$ for   $N = 1,0$,
which  corresponds to
the $\hats$ dependence of the $|$CI$|^2$
 and interference contributions respectively. For 
$ij = uu,dd$ or $ud$, these
integrals can be up to 20\% larger  than
for $N=-1$. We conservatively neglect this effect,
and use the $r_{ij}$ given in table \ref{tab:rij}.
\end{itemize}

\begin{table}[t]
\begin{center}
\begin{tabular}{|c|c|}
\hline
incoming partons $ij$ & $r_{ij}=$ integrated density \\
&(normalised to $uu$)\\
 \hline
$ u~u$ & 1\\
$d~d$&  0.085\\
$ u ~ d$& 0.56\\
$ u ~ g$& 0.22\\
\hline
\end{tabular}
\caption{ The ratios  of integrated pdfs, $r_{ij}= I^{-1}_{ij}/I^{-1}_{uu}$, 
which allow translating
the experimental bound on  $\sigma(pp\to $ dijets) 
to a bound on partonic cross sections. $r_{ud}$ and  $r_{ug}$ are multiplied by 2 
because the $u$ {valence quark} could {come from} either incoming proton. 
\label{tab:rij}}
\end{center}
\end{table}

{In the end,}
as desired, we have obtained an analytic formulation of the
experimental bound on contact interactions. The data gives
$\epsilon \lsim  1/3$ or 1/2. For a contact interaction selected from
section~\ref{sec:inteff}, with incoming flavours $ij$,
\beq
\epsilon = \frac{r_{ij} \delta \frac{d\hat{\sigma}}{d\chi} 
(q_i q_j \to  q_m q_n)}
{(1 + r_{dd}) \frac{d\hat{\sigma}_{QCD} }{d\chi}
(uu \to uu ) +  r_{ud} \frac{d \hat{\sigma}_{QCD} }{d\chi} 
(ud \to ud )
+ r_{ug} \frac{d\hat{\sigma}_{QCD} }{d\chi} 
(ug \to ug )
}
\label{bound}
\eeq
where $r_{ij}$ is from table  \ref{tab:rij},
$ \delta \frac{d \hat{\sigma}}{d\chi} $ is the parton-level
excess with respect to QCD induced by contact interactions ({given by}
the Appendix), whereas the cross sections in the denominator
 are induced by QCD (see eqs.~(\ref{P+S17.71}), (\ref{P+S17.70}) and (\ref{P+S:17.64})).

\begin{table}
\begin{center}
\begin{tabular}{|l|c|c|c|}
\hline
Operator&flavours $minj$ & $\hats/\Lambda^2 $ &$\Lambda <$  \\
\hline
\hline
$+{\cal O}^8_{XY}$ & $\overline{u}u\overline{u}u$ &.64~~ .72& $3.8 -3.5$  TeV\\
$-{\cal O}^8_{XY}$ & $\overline{u}u\overline{u}u$ & .20~~.28 & $6.7-5.7 $  TeV\\
&&&\\
$+{\cal O}^1_{XX}$ & $\overline{u}u\overline{u}u$ &.19~~.21 &$6.9-6.5$  TeV\\
$-{\cal O}^1_{XX}$ & $\overline{u}u\overline{u}u$ &.06~~.09& $12.2 -10$  TeV\\
$+{\cal O}^1_{XY}$ & $\overline{u}u\overline{u}u$ & .19~~.23& $6.9-6.3 $  TeV\\
$-{\cal O}^1_{XY}$ & $\overline{u}u\overline{u}u$ &.15~~.19& $7.7 -6.9$  TeV\\
\hline
%$+{\cal O}^8_{XY}$ & $\overline{d}d\overline{d}d$ & - & $ - $   \\
%$-{\cal O}^8_{XY}$ & $\overline{d}d\overline{d}d$ & -& $- $  \\
%%
%&&&\\
$+{\cal O}^1_{XX}$ & $\overline{d}d\overline{d}d$ &.43 ~~.52& $4.6-4.2$  TeV\\
$-{\cal O}^1_{XX}$ & $\overline{d}d\overline{d}d$ &.31~~.39& $5.4-4.8 $  TeV\\
$+{\cal O}^1_{XY}$ & $\overline{d}d\overline{d}d$ &.60~~.74& $3.9- 3.5 $  TeV\\
$-{\cal O}^1_{XY}$ & $\overline{d}d\overline{d}d$ &.56~~.70& $ 4.0-3.6$  TeV\\
\hline
%
%&&&\\
$+{\cal O}^{pythia}_{XX}$ &  & .15~~.17& $7.7-7.3 $ TeV \\
 &&  & ($8.4$ TeV)\\% (CMS~\cite{CMS1}) \\
$-{\cal O}^{pythia}_{XX}$ & & .06~~.07 & $ 12.2-11.3$  TeV\\
& &  & ($11.7$ TeV)\\% (CMS~\cite{CMS1}) \\
$+{\cal O}^{pythia}_{XY}$ &  &.17~~.20& $7.3-6.7 $ TeV \\
& &  & ($8.0$ TeV)\\% (CMS~\cite{CMS1}) \\
$-{\cal O}^{pythia}_{XY}$ &  &.13~~.17& $8.3-7.3$ TeV \\
& &  & ($8.0$ TeV)\\% (CMS~\cite{CMS1}) \\
\hline
\end{tabular}
\caption{Estimated bounds on the contact interaction
scale, obtained from figure \ref{fig:uuuu}.
The interactions of the first column are 
from section~\ref{sec:inteff}, and
$ {\cal O}^{pythia}$ is the flavour-summed operator of
eq.~(\ref{inpy8}) for comparison.
The flavour
indices of the second column are in the order of the fields in
the operator, and correspond to $ij\to mn$. 
The bounds are
for $\alpha_s(M_{dijet}) = 0.09$. In the third
column are given the bounds on $\hat{s}/\Lambda^2$
from requiring that the relative excess
of  dijets induced by contact interactions
be   $|\epsilon|< 1/3$ or $|\epsilon|< 1/2$.
The bound in the last
column is obtained with
$ \hats = M_{dijet}^2 =( 3$ TeV$)^2$,
for the two values of $\epsilon$.
The bounds in parentheses 
 on $ {\cal O}^{pythia}$ are those of CMS~\cite{CMS1}.
\label{tab:qqqq}}
\end{center}
\end{table}

\begin{table}
\begin{center}
\begin{tabular}{|l|c|c|c|}
\hline
Operator&flavours $minj$ & $\hats/\Lambda^2 $ &$\Lambda <$  \\
\hline
\hline
$+{\cal O}^8_{XX}$ & $\overline{d}d\overline{u}u$ &.59~~.67& $ 3.9-3.7$  TeV\\
$-{\cal O}^8_{XX}$ & $\overline{d}d\overline{u}u$ &.21~~.28& $6.4-5.7 $  TeV\\
%$+{\cal O}^8_{XY}$ & $\overline{d}d\overline{u}u$ &-& $- $ \\
$-{\cal O}^8_{XY}$ & $\overline{d}d\overline{u}u$ &.50~~.66& $4.1-3.7 $  TeV\\
$~~{\cal O}^{S8}$ & $\overline{d}d\overline{u}u$ &..60~~.74& $3.9 - 3.5 $  TeV\\
&&&\\
%\hline
%\hline
$+{\cal O}^1_{XX}$ & $\overline{d}d\overline{u}u$ &.17~~.22 &$7.1-6.4$  TeV\\
$-{\cal O}^1_{XX}$ & $\overline{d}d\overline{u}u$ &.16 ~~.20& $7.3-6.7  $  TeV\\
$+{\cal O}^{1}_{CC}$ & $\overline{u}d\overline{d}u$ &.27 ~~.31& $5.8-  5.4$  TeV\\
$-{\cal O}^{1}_{CC}$ & $\overline{u}d\overline{d}u$ &.10~~.14 &$9.5-8.0$  TeV\\
$+{\cal O}^1_{XY}$ & $\overline{d}d\overline{u}u$ & .33~~.41& $5.1-4.7 $  TeV\\
$-{\cal O}^1_{XY}$ & $\overline{d}d\overline{u}u$ &.31~~.39& $5.2-4.8 $  TeV\\
$~~{\cal O}^{S1}$ & $\overline{d}d\overline{u}u$ &.39~~.48& $4.8-4.3 $  TeV\\

\hline
\end{tabular}
\caption{ Estimated bounds on the contact interaction
scale, obtained from figure
\ref{fig:uudd} for the interactions of the first column 
with $ij \to mn = ud \to ud$ flavour structure.
 In the third
column are given the bounds on $\hat{s}/\Lambda^2$
from requiring    $|\epsilon| < 1/3$ or $|\epsilon| < 1/2$.
The last
column is obtained with
$ \hats = M_{dijet}^2 = (3$ TeV$)^2$.
\label{tab:qqpqqp}}
\end{center}
\end{table}

\begin{table}
\begin{center}
\begin{tabular}{|l|c|c|c|}
\hline
Operator&flavours $minj$ &  $\hats/\Lambda^2 $&$\Lambda <$  \\
\hline
${\cal O}^8_{XY}$ & $\overline{c}u\overline{c}u$ &.35~~.44 & $5.1-4.5 $  TeV\\
${\cal O}^1_{XX}$ & $\overline{c}u\overline{c}u$ &.11~~.13& $9.0-8.3 $  TeV\\
${\cal O}^1_{XY}$ & $\overline{c}u\overline{c}u$ &.17~~.21& $7.3-6.5 $  TeV\\
&&&\\
${\cal O}^1_{XX}$ & $\overline{s}d\overline{s}d,
\overline{b}d\overline{b}d$ &.37~~.45& $ 4.9- 4.5$  TeV\\
${\cal O}^1_{XY}$ & $\overline{s}d\overline{s}d,
\overline{b}d\overline{b}d$ &.59~~.71& $3.9-3.6  $  TeV\\
%
%${\cal O}^8_{XY}$ & $\overline{s}d\overline{b}d$ &- & $- $  \\
${\cal O}^1_{XX}$ & $\overline{s}d\overline{b}d$ &.30 ~~.37& $5.5-4.9  $  TeV\\
${\cal O}^1_{XY}$ & $\overline{s}d\overline{b}d$ &.59~~.71& $3.9-3.6  $  TeV\\
\hline
\end{tabular}
\caption{ 
 Estimated bounds on the contact interaction
scale, obtained from figure
\ref{fig:uucc}, for the interactions of the first column 
with $ij \to mn$ flavour structure,
$ij = uu$ or $dd$, and $\Delta F=2$.
% (see the caption of figure \ref{tab:qqqq}).
 In the third
column are given the bounds on $\hat{s}/\Lambda^2$
from requiring   $|\epsilon| < 1/3$ or $|\epsilon| < 1/2$.
The bound in the last
column is obtained with
$ \hats = M_{dijet}^2 = (3$ TeV$)^2$.
\label{tab:DF4NC}}
\end{center}
\end{table}

\begin{table}
\begin{center}
\begin{tabular}{|l|c|c|c|}
\hline
Operator&flavours $minj$ &  $\hats/\Lambda^2 $&$\Lambda <$  \\
\hline
${\cal O}^8_{XY}$ & $\overline{c}u\overline{u}u,
\overline{u}u\overline{c}u$ 
&.35~~.44 & $5.1-4.5 $  TeV\\
${\cal O}^1_{XX}$ & $\overline{c}u\overline{u}u$ 
&.09~~.11& $10-9.0 $  TeV\\
${\cal O}^1_{XY}$ & $\overline{c}u\overline{u}u,
\overline{u}u\overline{c}u$ 
&.16~~.21& $7.5-6.5 $  TeV\\
&&&\\
${\cal O}^1_{XX}$ & $\overline{q}d\overline{d}d
%\overline{d}d\overline{q}d
$ &.30~~.37& $ 5.5-4.9   $  TeV\\
${\cal O}^1_{XY}$ & $\overline{q}d\overline{d}d,
\overline{d}d\overline{q}d$ 
&.59~~.71& $3.9-3.6  $  TeV\\
\hline
\end{tabular}
\caption{
 Estimated bounds on the contact interaction
scale from figure \ref{fig:uucc}, 
for the interactions of the first column 
with flavour structure $ij \to mn$,
$ij = uu$ or $dd$, and $\Delta F=1$.
In this table,  $q = b,s$.
 In the third
column are given the bounds on $\hat{s}/\Lambda^2$
from requiring   $|\epsilon| < 1/3$ or $|\epsilon| < 1/2$.
The bound in the last
column is obtained with
$ \hats = M_{dijet}^2 = (3$ TeV$)^2$.
\label{tab:DF2NC}}
\end{center}
\end{table}

\begin{table}
\begin{center}
\begin{tabular}{|l|c|c|c|}
\hline
Operator& flavours $minj$ &  $\hats/\Lambda^2 $&$\Lambda <$  \\
\hline
${\cal O}^8_{XX}$ & $\overline{c}u\overline{q}d $ &.36~~.43 & $5.0-4.6 $  TeV\\
${\cal O}^8_{XY}$ & $\overline{c}u\overline{q}d,  \overline{q}d\overline{c}u $ 
&.67~~.84 & $3.7-3.3 $  TeV\\
${\cal O}^1_{XX}$,${\cal O}^1_{CC}$ 
& $\overline{c}u\overline{q}d$ &.20~~.25& $6.7-6.0 $  TeV\\
%${\cal O}^1_{CC}$ & $\overline{q}d\overline{c}u$ &.20~~.25& $6.7-6.0 $  TeV\\
${\cal O}^1_{XY}$ & $\overline{c}u\overline{q}d,\overline{q}d\overline{c}u $ 
&.32~~.40& $5.3-4.7 $  TeV\\
${\cal O}^{S1}$ & $-\overline{c}u\overline{q}d+\overline{q}u\overline{c}d $ 
&.56~~.67& $4.0-3.7 $  TeV\\
\hline
\end{tabular}
\caption{ 
 Estimated bounds on the contact interaction
scale from figure \ref{fig:uucc}, for the interactions of the first column 
with flavour structure $ij \to mn$,
$ij = ud$, and $\Delta F=2$.
In this table, $q = b,s$.
 In the third
column are given the bounds on $\hat{s}/\Lambda^2$
from requiring   $|\epsilon| < 1/3$ or $|\epsilon| < 1/2$.
The bound in the last
column is obtained with
$ \hats = M_{dijet}^2 =( 3$ TeV$)^2$.
\label{tab:DF4CC}}
\end{center}
\end{table}

\begin{table}
\begin{center}
\begin{tabular}{|l|c|c|c|}
\hline
Operator& flavours $minj$ &  $\hats/\Lambda^2 $&$\Lambda <$  \\
\hline
${\cal O}^8_{XX}$ & $\overline{c}u\overline{d}d ,
\overline{u}u\overline{q}d $ &.36~~.43 & $5.0-4.6 $  TeV\\
${\cal O}^8_{XY}$ & $\overline{c}u\overline{d}d,  
\overline{d}d\overline{c}u $ 
&&\\
& $\overline{u}u\overline{q}d,  
\overline{q}d\overline{u}u $ 
&.66~~.84 & $3.7-3.3 $  TeV\\
${\cal O}^1_{XX}$,${\cal O}^1_{CC}$ 
& $\overline{u}u\overline{q}d$, $\overline{c}u\overline{d}d$ &.20~~.25& $6.7-6.0 $  TeV\\
${\cal O}^1_{XY}$ & $\overline{c}u\overline{d}d,\overline{d}d\overline{c}u $ 
& &  \\
 & $\overline{u}u\overline{q}d,\overline{q}d\overline{u}u $ 
&.32~~.40& $5.3-4.7 $  TeV\\
${\cal O}^{S1}$ & $-\overline{c}u\overline{d}d+\overline{d}u\overline{c}d $ 
& &\\
 & $-\overline{u}u\overline{q}d+\overline{q}u\overline{u}d $ 
&.56~~.67& $4.0-3.7 $  TeV\\
\hline
\end{tabular}
\caption{
 Estimated bounds on the contact interaction
scale from figure \ref{fig:uucc}, for the interactions of the first column 
with flavour structure $ij \to mn$, $ij = ud$, and $\Delta F=1$.
$q$ = $b$ or $s$. In the third
column are given the bounds on $\hat{s}/\Lambda^2$
from requiring    $\epsilon < 1/3$ or $\epsilon < 1/2$.
The bound in the last
column is obtained with
$ \hats = M_{dijet}^2 = ( 3$ TeV)$^2$.
\label{tab:DF2CC}}
\end{center}
\end{table}

\subsection{Bounds on flavoured operators}

{We are now in a position to translate the CMS results in terms of flavoured operators.} The recipe to guess a bound on $\Lambda$, as given {in the previous section}, can now be reformulated
analytically:
\begin{enumerate}
\item  look up in the Appendix 
the contribution to the 
partonic cross section  of  the selected  
contact interaction(s) plus interference with  QCD,
and evaluate it at the $\chi$  corresponding to the bin of interest~\footnote{For instance, $1\leq \chi \leq 3$ corresponds to the range
in $(-\hatt,-\hatu)$ between $(\frac{3}{4}\hats,\frac{1}{4}\hats)$ and 
$(\frac{1}{4}\hats,\frac{3}{4}\hats)$.}. {The same must be done} for  the QCD cross sections
(\ref{P+S17.71}), (\ref{P+S17.70}) and (\ref{P+S:17.64}).

\item  weight the various contact interactions by
the appropriate $r_{ij}$ factor from table \ref{tab:rij},
and the QCD cross sections as given in the
denominator of eq.~(\ref{bound}).

\item  impose that the ratio  of eq.~(\ref{bound}),  $\epsilon \leq 1/3$ (or 1/2),
which gives a quadratic polynomial for 
$\hats/\Lambda^2$ whose root gives 
the estimated bound on $\Lambda$ for
$\hats$ taken at the {lower end of the range allowed in the}
highest dijet mass bin.

\end{enumerate}

\begin{figure}[h!]
\begin{center}
\epsfig{file=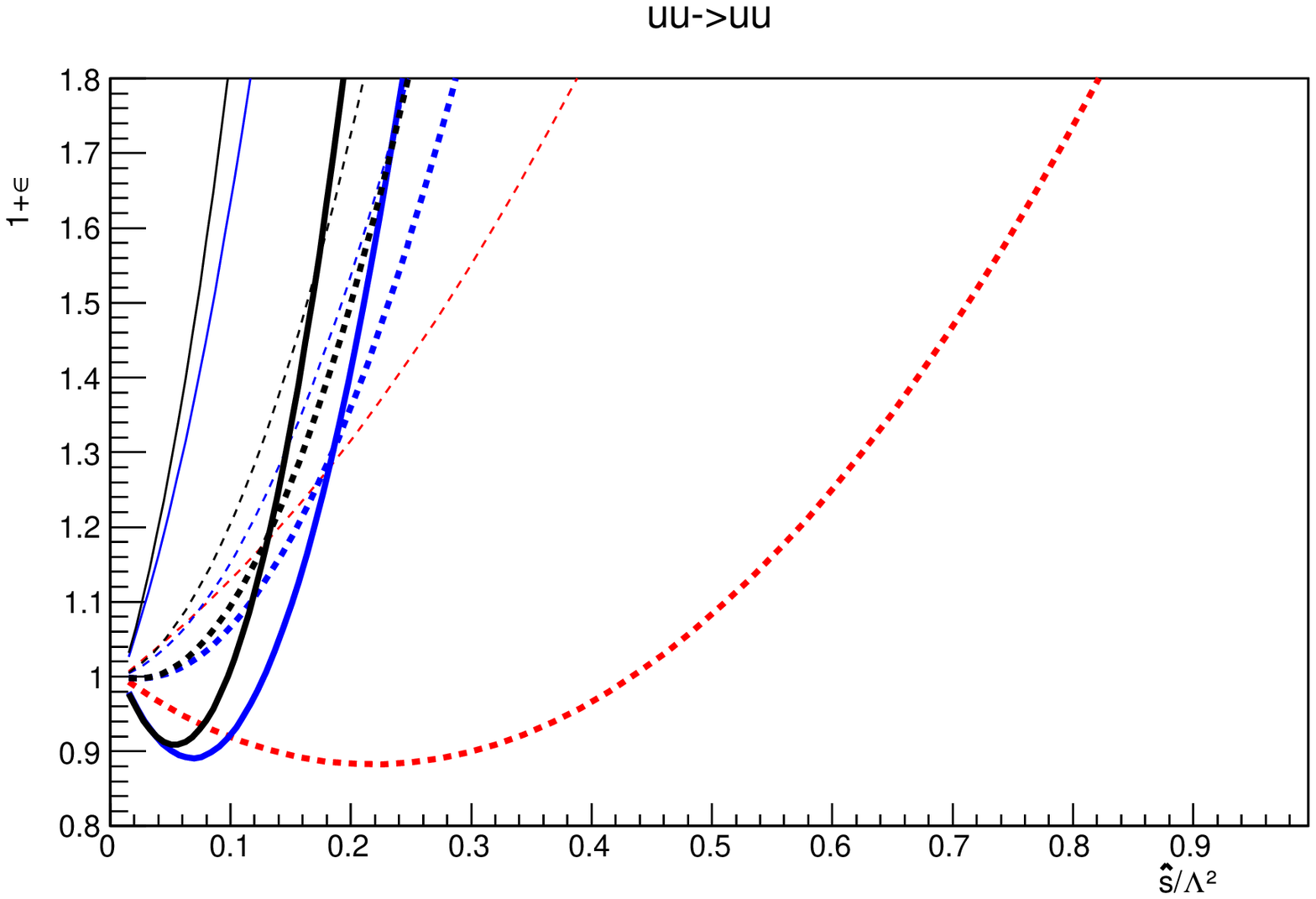, width=15cm}
\epsfig{file=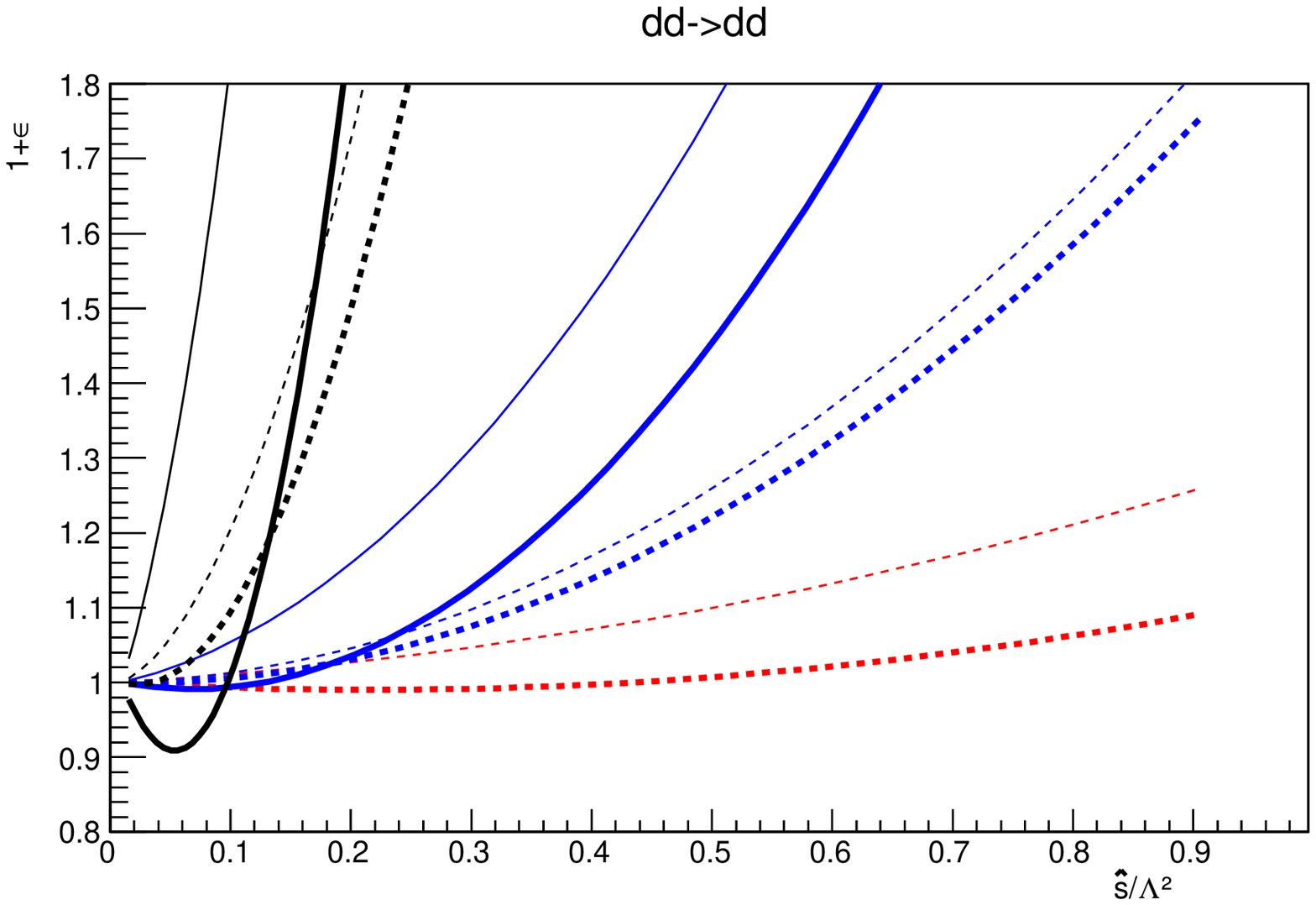, width=15cm}
\end{center}
%\vspace{-10mm}
\caption{{Ratio} $1+\epsilon$ (see eq.~(\ref{bound})) as a function of 
$\hats/\Lambda^2$ for  $uu\to uu$ (up) and  $dd\to dd$ (down) contact 
interactions.  Continuous (dashed) lines correspond to  
interactions with the same (opposite)  chiral projector  
in the two  currents, and  thick (thin) lines are 
for a positive (negative) coefficient  in the Lagrangian. 
Red lines   are for ${\cal O}^8$,  and blue for  ${\cal O}^1$. 
Black are  the flavour-summed ${\cal O}^{pythia}$ 
for  comparison. {The bounds derived on $\Lambda$ are obtained by
requesting that $\hats/\Lambda^2$ should}
be small enough to give $1+\epsilon \lsim 1.33 \to 1.5$.   }
\label{fig:uuuu}
\end{figure}

\begin{figure}[h!]
\begin{center}
\epsfig{file=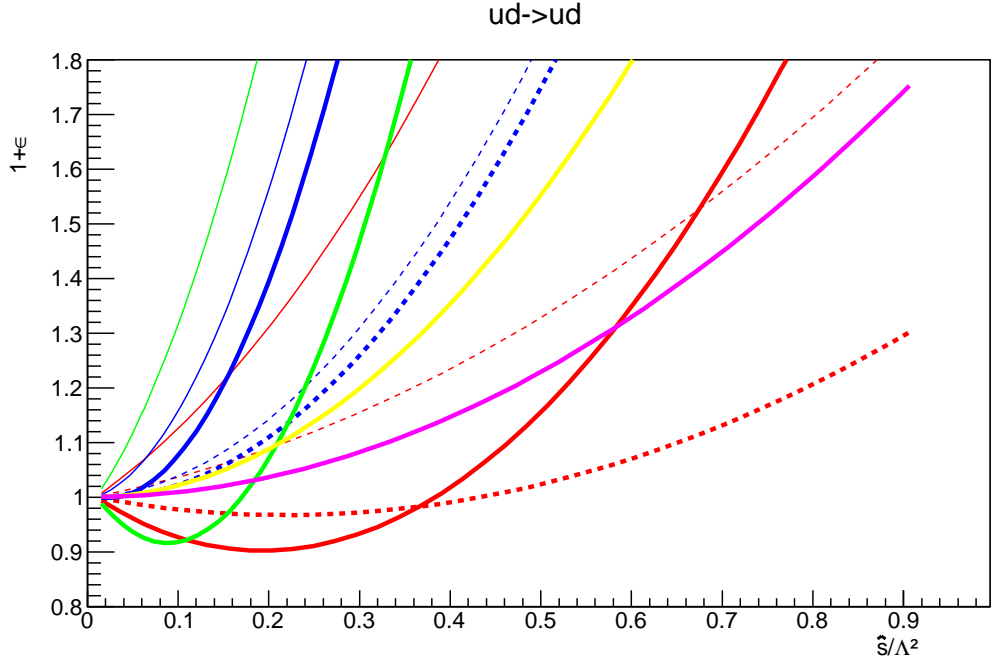, width=15cm}
\end{center}
%\vspace{-10mm}
\caption{{Ratio} $1+\epsilon$ (see eq.~(\ref{bound})) as a function of $\hats/\Lambda^2$ for  $ud\to ud$ contact interactions.  Continuous (dashed) lines correspond to  interactions with the same (opposite)  chiral projector  in the two  currents, and  thick (thin) lines are for a positive (negative) coefficient  in the Lagrangian. Red lines are for ${\cal O}^8$, blue for  ${\cal O}^1$, green for  ${\cal O}^1_{CC}$, {yellow for ${\cal O}^{S1}$, pink for ${\cal O}^{S8}$}.}
\label{fig:uudd}
\end{figure}

\begin{figure}[h!]
\begin{center}
\epsfig{file=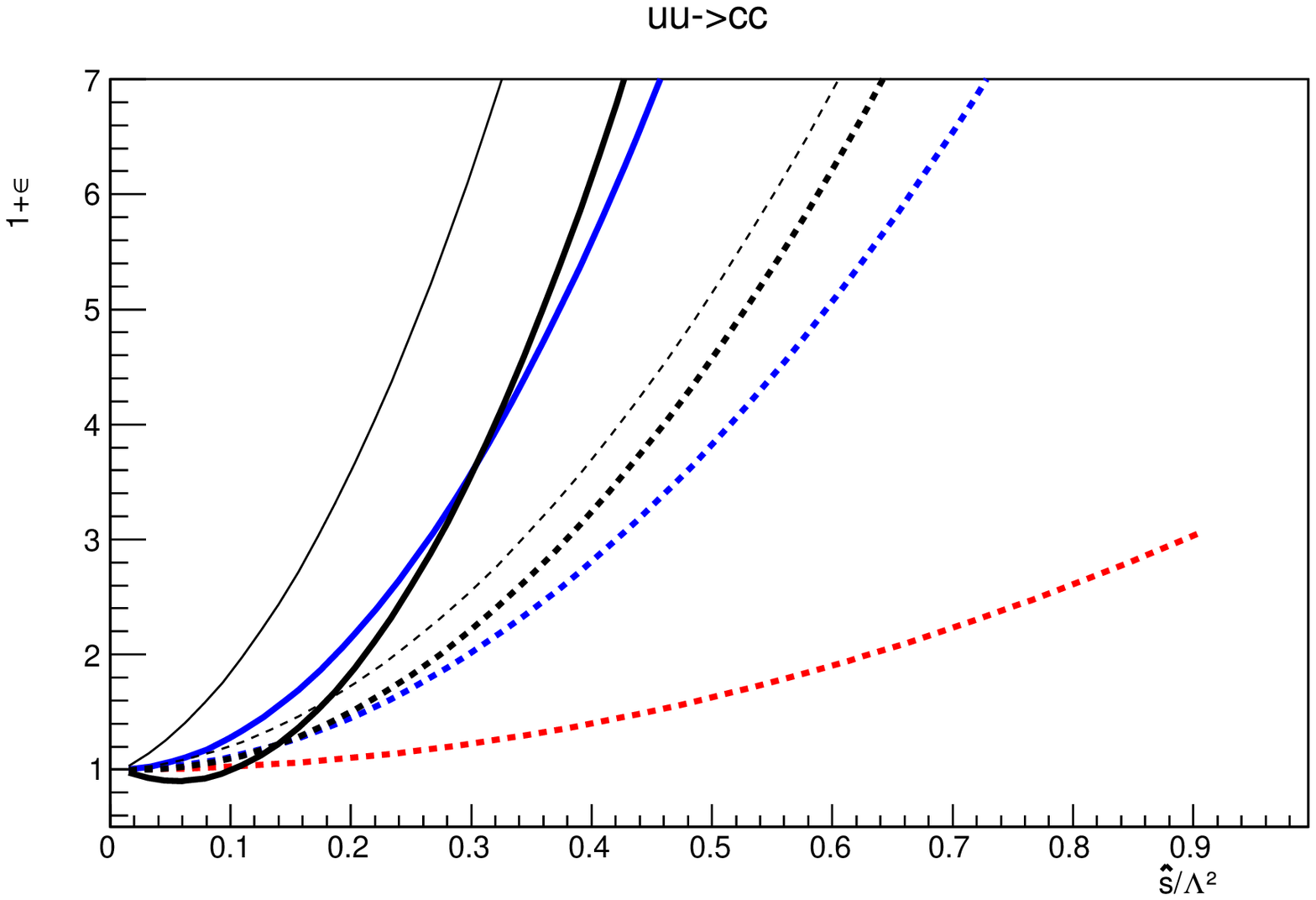, width=15cm}
\epsfig{file=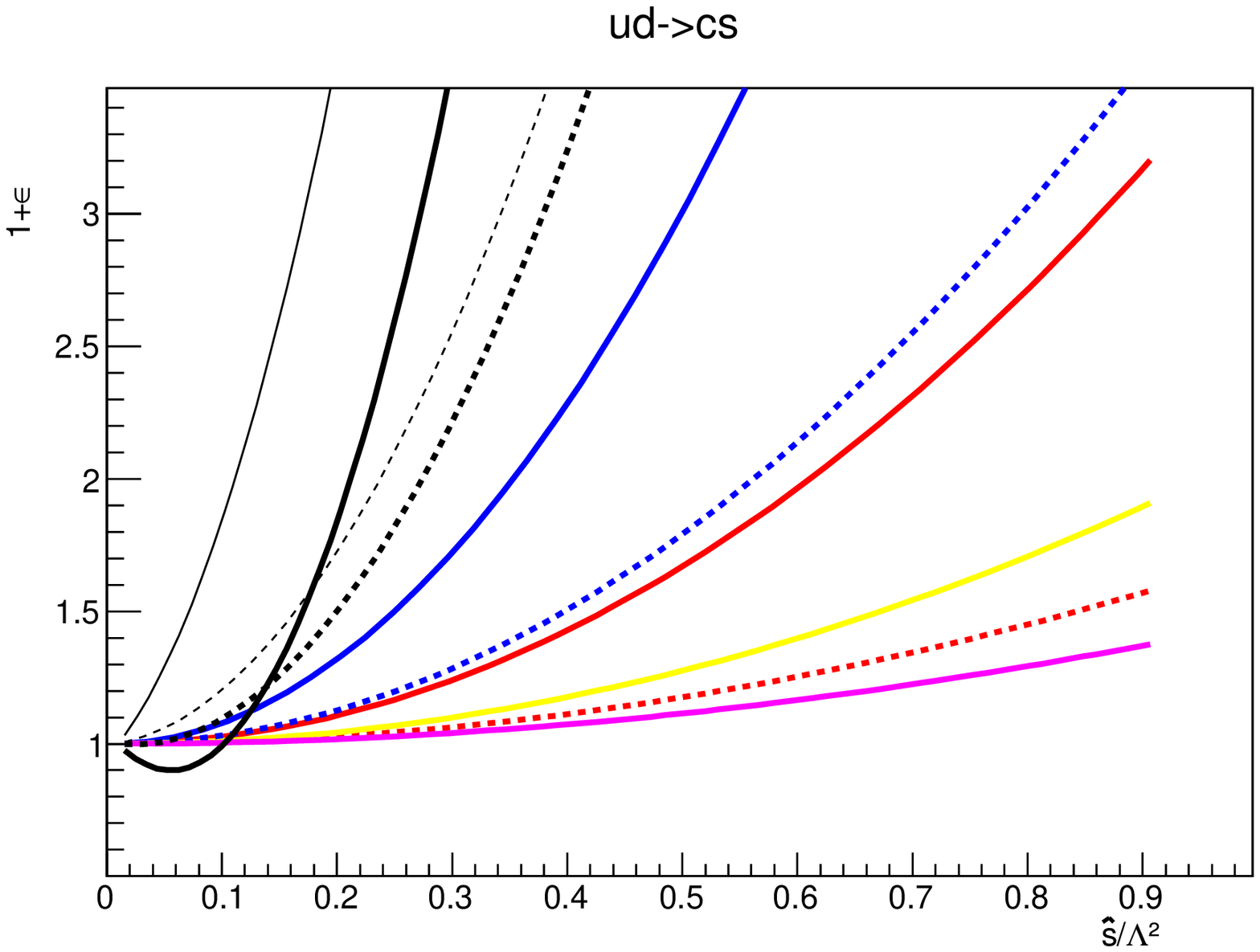, width=15cm}
\end{center}
%\vspace{-10mm}
\caption{{Ratio} $1+\epsilon$ (see eq.~(\ref{bound})) as a function of $\hats/\Lambda^2$ for flavour changing contact interactions
 $uu \to cc$ (up) or $ud \to cs$ (down). 
Continuous (dashed) lines correspond to  interactions with the same 
(opposite)  chiral projector  in the two  currents.
There is no interference with QCD, so the
sign of the contact interactions does not matter.
 Red lines are for ${\cal O}^8$, blue for  ${\cal O}^1$, 
{yellow for ${\cal O}^{S1}$, pink for ${\cal O}^{S8}$}. 
Black are  the flavour-summed ${\cal O}^{pythia}$ for comparison.}
\label{fig:uucc}
\end{figure}

 In practice,
we integrate 
the partonic cross sections 
over $1\leq \chi\leq 3$,  and plot
$1+\epsilon$ in figures \ref{fig:uuuu},
\ref{fig:uudd} and  \ref{fig:uucc}.
Imposing
$\epsilon\leq 1/3$ or $1/2$ gives the
bounds on $\Lambda$ in the tables \ref{tab:qqqq} to \ref{tab:DF2CC}, {stemming from}
the ratio $M_{dijet}^2/\Lambda^2$,
by taking $M_{dijet} = 3$ TeV~\cite{CMS1}.

The bound {that} we obtain on $\Lambda$ is the solution
of a quadratic polynomial in $\hats/(\alpha_s\Lambda^2)$
(see eq.~(\ref{soln})),
so depends on the numerical value of $\alpha_s$. We
take $\alpha_s(3{\rm TeV}) \simeq 0.09$, in agreement with
the leading order running, since we do a leading order
calculation (this is analogous to using
 $\alpha_s (m_Z) = .139$  in {\sc pythia}~\cite{peter}). 
If instead, we take the Particle Data Group  value~\cite{PDG}
 $\alpha_s (m_Z) \simeq .12$, then 
 $\alpha_s(3{\rm TeV}) \simeq 0.08$. 
If the scale of evaluation of $\alpha_s$
is changed  by a factor of two, $\alpha_s$
varies by about 0.005. We conclude that
varying   $\alpha_s$ between  0.08 and 0.09 
gives some notion of the NLO uncertainties,
and take the larger value, since this yields 
the more conservative limit.

The study performed here neglects several {effects}, such as
 hadronisation (partons are not jets), and
NLO corrections (calculated for  ${\cal O}^{1,LL}$ in ref.~\cite{NLO}).
We also do not consider 
 dimension 8 operators:
as pointed out by ref.~\cite{CCL}, matrix elements to which QCD contributes
could also have an ${\cal O} (1/\Lambda^4)$ term from
QCD interference with a dimension 8 operator.
This   interference  is in principle suppressed
with respect to   dimension 6-{contributions} by a factor $\alpha_s$.

We do include interference between QED and contact interactions, when there
is no interference with QCD, but we neglect {effects of weak interactions}.
There are (almost) no gluon contributions in this analysis: we
include the $gq\to gq$ contribution to
the QCD cross section, but
neglect $gg\to gg$ (because we
assume $f_g \lsim f_d/3\lsim f_u/9$).
We hope to  consider $gg\bar{q} q$
contact interactions~\cite{ggqq} in a later analysis.

\subsection{Extrapolating to higher energy or luminosity}
\label{sec:slumi}

The reach of the future LHC for the
usual flavour-summed  contact interactions
 has been studied in ref.~\cite{snowmass}, who
find expected limits $\Lambda \gsim 20$ TeV. 
If the LHC with more energy and luminosity still
does not find contact interactions, how would our
flavoured estimates scale ?
The bounds obtained here are on the dimensionless variable
$\hats/\Lambda^2$, and they depend on the
experimental uncertainty via $\epsilon$, as well
as the scale at which the pdfs were evaluated.
There are two useful approximations/assumptions:
\ben
\item  suppose that the  ratios of integrated pdfs, $r_{ij}$, 
given in table
\ref{tab:rij}, will not change significantly in going
from the  8 to 14 TeV LHC. 
\item  suppose the last bin in dijet mass will always
have an experimental uncertainty of 20\%-30\% {and remain compatible with the QCD prediction}. This
may be reasonable,  because the bound on contact
interactions profits more from going to a larger
$\hats$ than from reducing the statistical uncertainties
to the size of the systematics. 
So $\epsilon$ would remain approximately 1/3 (or 1/2).
\een
Then the estimated bounds we quote on $\hats/\Lambda^2$
remain valid, and the bound on $\Lambda$ will be
multiplied by a factor $M_{dijet}^{(new)}$/(3 TeV), where
 $M_{dijet}^{(new)}$ is the  lower bound on the dijet mass of the 
highest bin of a future analysis.

In relation with this issue, we come back to the interpretation of the current result
of CMS in the lowest-$\chi$ bin  for $M_{dijet}>3$ TeV~\cite{CMS1}. 
Up to now, we have considered that
the spread of data below and above the  QCD predictions precluded 
explaining with contact interactions
 the $\sim 30\%$ deficit
in the $1\leq \chi \leq 3$ bin. Let us however
entertain this possibility, which requires a negative contribution coming from a negative interference between QCD and contact interactions. As already discussed earlier and explicitly seen from figures \ref{fig:uuuu},
\ref{fig:uudd} and  \ref{fig:uucc}, and in the Appendix, the sign and size of the interference depend on the flavour considered. It turns out that one can
reduce the number of events by around 10\% in the case of  $O^{1,XX}_{uuuu}$,
 $O^{1,CC}_{uddu}$,
$O^{8,XY}_{uuuu}$ and $O^{8,XX}_{uddu}$ (with a scale of around 11 TeV for the singlet operators
and a scale around 7 TeV for the octet operators).
These  operators show a similar behaviour once
extrapolated to higher $M_{dijet}$. They yield a deficit of 
events around 10\% in the bin 2.4-3 TeV 
(where CMS data are in very good agreement with QCD), 
and of around 7\% in the bin 1.9-2.4 TeV. 
The effect is within QCD uncertainties at lower $M_{dijet}$.

At higher $M_{dijet}$, of interest in the context of the LHC upgraded to 14 TeV, the contribution from
contact interactions-squared becomes larger and starts 
dominating over the interference with QCD:
for  4 TeV, the  total contribution from contact interactions 
 approximately vanishes,
and becomes positive  at higher dijet mass, with an excess of 30\% at
5 TeV, and more than 100\% at 6 TeV. In this particular scenario, the slight deficit  currently observed  in the $M_{dijet}> 3$ TeV bin is not enough 
to draw definite conclusions on the presence or 
absence of contact interactions, and this situation will still 
hold with the increase of the dijet energy, 
up to the point where one reaches so high dijet masses that
the (now positive) contribution from contact interactions yields a significant, non ambiguous, sign of their presence.

On the other hand, going to such high dijet masses, one could in principle discriminate between singlet and octet cases, since the latter case
corresponds to a scale $\Lambda\sim 6$ TeV and should thus be resolved in terms of intermediate massive states, whereas the former, linked with a scale 
$\Lambda\sim $ 11 TeV, could still remain a contact interaction 
if studied by a LHC running at 14 TeV.

%%%%%%%%%%%%%%%%%%%%%%%%%%%%%%%%%%%%%%%%%%%%%%%%%%%%%%%%%%%%%%%%%%%%%%
%% SECT  %%%%%%%%%%%%%%%%%%%%%%%%%%%%%%%%%%%%%%%%%%%%%%%%%%%%%%
%%%%%%%%%%%%%%%%%%%%%%%%%%%%%%%%%%%%%%%%%%%%%%%%%%%%%%%%%%%%%%%%%%%%%%

\section{Discussion}
\label{sec:disc}

Contact interactions can be induced by the exchange of 
new  resonances  which are not resolved as
mass eigenstates,  because they are 
either broad or heavier than the exchanged
four-momentum. 
The  tree-level exchange of an
off-shell boson of mass $M_{new} \gsim M_{dijet}$, 
interacting with the quarks via a coupling
$g'$, would give  $4\pi/\Lambda^2 \sim g^{'2}/M^2$.
However, the order of magnitude of collider
bounds,  estimated in eq.~(\ref{ordgran}),
is  $1/\Lambda^2 \sim \alpha_s/\hats$,
so $M_{new}^2 \gsim M_{dijet}^2$ if $\alpha' \gsim \alpha$.
The contact interaction approximation
is appropriate at a collider 
to describe  the exchange of particles with
${\cal O}(1)$ couplings.   So if the 14 TeV LHC reaches
a sensitivity  of $\Lambda\sim 20 - 30 $ TeV, this excludes
strongly coupled particles,  without a  conserved
parity, up to masses $\sim 10 $ TeV.

Heavy particles interacting with quarks can 
evade contributing to contact interactions in
various ways. If new particles have a conserved
parity (as is convenient  to obtain dark matter),
and couplings $g' \lsim g_s$,
they can   generate contact interactions
via a closed loop of heavy new particles. This
implies {either} a contact interaction  coefficient 
\begin{equation}
4\pi/\Lambda^2 \sim g^{'4}/(16\pi^2 M_{new}^2)~,
\end{equation}
which is small enough to be allowed for $M_{new} \sim M_{dijet}$
(it gives $\Lambda >2 M_{new}/\alpha'$), or {the absence of}
a contact interaction for 
\begin{equation}
g'^{4}/(16\pi^2 M_{new}^2) \sim 
g_s^{2}/M_{dijet}^2~,
\end{equation}
since the new particles
could then be {produced in pairs}. 
Heavy new bosons that are less strongly coupled
to quarks can also evade contact interaction
bounds. For instance, a $Z'$ with Standard Model
couplings induces various four-quark contact interactions
with a coefficient 
 $ 4\pi/\Lambda^2 \sim g^{2}/(8 c_W^2 M_{Z'}^2)$.

Despite that a contact interaction which is
absent at 3 TeV might be present  at lower
energies, it is interesting to compare
the collider  bounds on
four-quark contact interactions
 to
limits from precision flavour physics.
The first step should be to 
evolve the operator coefficients between the TeV scale 
and   low energy
($e.g.~m_b$):
$$
\left. \frac{4 \pi}{\Lambda^2}\right|_{m_b} \simeq
c \left.  \frac{4 \pi}{\Lambda^2}\right|_{3~ {\rm TeV}} 
\sim \left. \left( \frac{\alpha_s(3 ~ {\rm TeV})}{\alpha_s(m_b)}\right)^{\gamma/2\beta_0} \frac{4 \pi}{\Lambda^2}\right|_{3~ {\rm TeV}} 
$$
where $\beta_0= (11N_c -2N_f)/12\pi$. For a few cases 
where we know the anomalous
dimension $\gamma$ of the contact interaction,  $0.5 < c<1.5$~\cite{Buras,UTfit}, so we
neglect the running and use at low energy the bound 
on $4\pi/\Lambda^2$ obtained at 3 TeV. 
{Due to the importance of Fermi interaction in low-energy} precision physics,
 it is convenient to define {the parameter $\beta$ as}
$$
\frac{4 \pi}{\Lambda^2} = \beta\frac{4 G_F}{\sqrt{2}}
\Rightarrow \beta = \frac{4 \pi v^2}{\Lambda^2} =
\left(\frac{0.6~ {\rm TeV}}{\Lambda}\right)^2
$$ 
so the collider bounds {discussed previously, ranging from 3 to 11 TeV,} imply $\beta \lsim 4 \times 10^{-2} 
\to 3 \times 10^{-3}$. 
Let us add that flavour physics and collider searches do not have the same scope in probing the flavour structure of contact interactions. As discussed above, proton-proton collider searches have the potential to search four-quark contact interactions involving at least two quarks of the first generation, but the two other quark lines are left free (and perhaps could be identified through jet tagging). 
In flavour physics,  neutral-meson mixing  is a sensitive probe
of flavour  off-diagonal ($\Delta F =2$) contact interactions~\cite{UTfit,Flavourbounds},
with particularly stringent bounds on $\Lambda$ for operators 
inducing kaon mixing (assuming, as we have done here, that the coefficients $\eta$ are $O(1)$). Bounds on other four-quark operators 
could in principle come from processes where a meson decays into two lighter mesons non-leptonically -- however, such processes
are very challenging from the theoretical point of view, and can 
hardly be considered as useful to set constraints
on contact interactions.

If a signal for contact interactions was observed
at the LHC,  it could indicate
 strongly coupled New Physics  at  a scale
just beyond the  reach of the LHC, or
the  (perturbative) exchange of a (broad) resonance
in $t$ or $s$-channel. In either case, 
it would be  interesting to identify the
flavour of the final state jets --- not
only to distinguish  gluons and  heavy flavours ($b,c$)
from light {quarks},  but even
to distinguish  $u$ from $d$ jets using jet charge. 
This would allow one
to predict the expected rate for the crossed process.
For instance, if a $t$-channel $Z'$ mediates 
$uu\to uu$, then it could also be
searched for as a bump in $u \bar{u} \to u \bar{u}$ {whereas}
an $s$-channel  diquark inducing the contact
interactions $ud\to cs$ is not worth
searching for in $t$ channel.

\section{Summary}
\label{sec:summ}

Many models with new particles at high energy have contact interactions
as their low-energy footprints. As such, contact interactions
are a  parametrisation of New Physics, so it is important that
constraints  on upon them be as widely applicable as possible. 
Current collider bounds  are calculated for a palette
of colour-singlet interactions (no colour matrices
in the vertex), summed on flavour.
In this paper, we estimate bounds on an almost complete basis
of  four-quark contact interactions,  with specified
flavour indices mediating $q_i q_j \to q_m q_n$,
where  $q_i, q_j$ are first-generation quarks.

We start from a  basis of dimension
six, Standard Model gauge-invariant, four-quark operators.
After electroweak symmetry breaking, each
operator induces one or  several four-quark interactions.
We constrain
 the coefficients of these effective interactions
 by turning them on one
at a time. We prefer to constrain the coefficients
of the effective interactions, rather than the gauge-invariant
operators, because this allows us to apply the
bounds to a different operator basis.  The effective
interactions (almost) do not interfere
among themselves, so {that} the bounds obtained by
turning them on one at a time are
conservative. A more
stringent bound could apply in the presence of
several contact interactions.

The bounds profit from a  convenient choice
of variables made by the experimental collaborations, 
which allows one to approximate the $pp \to $ dijets
cross section as an integral over parton distribution
functions multiplying a partonic
cross section.  We estimate the expected limit
on contact interactions by comparing their partonic
cross sections to the leading order QCD prediction.
The data agree with  QCD, so we require
that the contact interactions   contribute  less than 1/3 or 1/2 
of the QCD expectation. 
Our bounds    are listed in tables \ref{tab:qqqq}-\ref{tab:DF2CC}.
Our flavoured estimates are generically  lower than the
limits of ATLAS and CMS, who constrain a flavour-sum  of contact
interactions given in eq.~(\ref{inpy8}). Our estimate
for this flavour-summed operator is comparable to the
experimental bounds.

The original aim of this paper was to
identify  ``classes'' among
the large collection of  contact interactions,
such that a bound obtained on a  representative 
of the class could be translated by
some simple rule to the others. The situation is however more complicated:
 the obstacle seems
to be the  interference between QCD and
the contact interaction, which precludes
any simple scaling of  the limit on $\hats/\Lambda^2$
from one interaction to another \footnote{ 
It is mildly surprising  that a limit
 on contact interactions at
a hadron collider can be estimated analytically;  however,
in practice, it may be just as simple to
use Madgraph5~\cite{mg}.}. 
An interesting feature of our analytic recipe
is that it constrains the dimensionless ratio
$M^2_{dijet}/\Lambda^2$ for each contact
interaction. 
The future LHC, with more energy
and luminosity, will be able to  probe a
higher dijet mass than the 
 $M_{dijet} = 3$ TeV~\cite{CMS1} used
here to extract  bounds on $\Lambda$.
The estimated bounds on  $\Lambda$ from
tables \ref{tab:qqqq}-\ref{tab:DF2CC}
should  therefore scale as $M_{dijet}(new)/(3$ TeV),
under assumptions given in
section~\ref{sec:slumi}.

Finally, we discussed briefly which contact interactions could explain 
a moderate deficit of events in the current data for the 
$1\leq \chi\leq 3$ and $M_{dijet}>3$ bin,
and how this would extrapolate to higher dijet masses.
It will  be particular interesting to see how the spread of results
obtained by the experimental collaborations evolve as luminosity and energy increase.

\section{Acknowledgements}

We thank  M. Gouzevitch,  A. Hinzmann, G. Salam, and  P. Skands for useful
conversations. This project was
performed in the context of the Lyon Institute of Origins,
ANR-10-LABX-66.

%%%%%%%%%%%%%%%%%%%%%%%%%%%%%%%%%%%%%%%%%%%%%%%%%%%%%%%%%%%%%%%%%%%%%%
%% SECT  %%%%%%%%%%%%%%%%%%%%%%%%%%%%%%%%%%%%%%%%%%%%%%%%%%%%%%
%%%%%%%%%%%%%%%%%%%%%%%%%%%%%%%%%%%%%%%%%%%%%%%%%%%%%%%%%%%%%%%%%%%%%%

\appendix 

\section{Kinematics and cross sections}
\label{app}

This Appendix gives the  contact interaction correction, 
in sections labeled by flavour structure, for a larger
set of contact interactions than the basis of 
section~\ref{sec:inteff} (in particular, we
give cross sections for octet
interactions involving four $u$- or $d$-type
quarks,although these are not in the list of
section~\ref{sec:inteff}).   The formulae are
labeled to the right by the contact interaction
to which they apply.
The partonic cross sections mediated by contact interactions
are given in several references~\cite{CCL,spira} 
\footnote{These formulae do not always agree
with~ref.~\cite{eichten}, and disagree on the sign of
some interferences with respect to ref.~\cite{lane}}, and  are collected below
 for convenience.

\subsection{Definitions}

 For momenta as
given in  the first diagram
of figure \ref{figQCDLam}, with time running left to right,
the Mandelstam variables are defined as
\bea
\hat{s} &=&  (k_1 + k_2)^2 = 2 k_1 \cdot k_2 
= 2 p_1 \cdot p_2  = 4 E^{*2}
 \nonumber\\
\hat{t} &=&  (k_1 - p_1)^2 = -2 k_1 \cdot p_1 = -2 p_2 \cdot k_2  
= -2 E^{*2}(1 - \cos \theta^*)
 \nonumber\\
\hat{u} &=&   (k_1 - p_2)^2 = -2 k_1 \cdot p_2 = -2 p_1 \cdot k_2  
= -2 E^{*2}(1 + \cos \theta^*)
\label{mandelstam}
\eea
where the last expressions are in the parton centre-of mass frame. If 
time runs upwards in the same diagram, 
 it describes $q \bar{q}$ annihilations,
and  the Mandelstam variables exchange their
definitions:
\bea
\hat{s}_{scat} =
 \hat{u}_{ann} % \nonumber\\
~~~,~~~
\hat{t}_{scat} = 
 \hat{s}_{ann} %  \nonumber\\
~~~,~~~
\hat{u}_{scat} 
= \hat{t}_{ann} 
\label{SDcrossingrules}
\eea 

In the following, we give  various
 partonic differential cross sections,  
$$
\frac{d \hat{\sigma}}{d \hatt} = \frac{\overline{|{\cal M}|}^2}{16 \pi \hats ^2}
\longrightarrow
\frac{d \hat{\sigma}}{d \chi} = \frac{\hatt^2}{\hats}
\left.\left(\frac{d \hat{\sigma}}{d \hatt} (\hats,\hatt,\hatu) +
\frac{d \hat{\sigma}}{d \hatt} (\hats,\hatu,\hatt)\right)
\right|_{\theta^*<\pi/2}
$$
where $\overline{|{\cal M}|}^2$ is averaged over incoming spins and colours,
and contains a factor 1/2 when the final state particles are identical. 
Since the change of variables to $\chi$ is simple
for cos$\theta^* >0$, we restrict  $0 \leq \theta^* \leq \pi/2$,
and add $d \hat{\sigma}/{d \hatt}$ with $\hatt \leftrightarrow \hatu$.

We include the QCD contribution, the contact-interaction interference with
QCD (and sometimes QED), and the contact interaction squared.
We neglect
the pure QED contribution (subdominant with respect to QCD), but include
QED-contact interference when there is no QCD-contact
interference. Indeed the interference term $\propto
\alpha_{em}\hats/\Lambda^2$ can
reduce  the cross section and have a minor effect on the bound.  

\subsection{$q_i g \to q_i g$ --- QCD only}

\bea
\frac{d\sigma}{d\hatt} = \frac{4 \pi \alpha_s^2 }{9\hats^2}
\left[
-\frac{\hatu}{\hats}
-\frac{\hats}{\hatu}
+\frac{9}{4}
\frac{\hats^2+ \hatu^2}{\hatt^2}\right]
\label{P+S17.70}
\eea

\subsection{$q_i q_j \to q_m q_n$}
\label{sigmaCI}

\subsubsection{$q q \to q q$,  ($ i = j = m = n$)}

\beq
 \frac{d \hat{\sigma}}{d \hatt}(qq \to qq) =
%\frac{1}{2}\left(
 \frac{2 \pi  \alpha_s^2 }{ 9 \hats ^2} 
\left[  \frac{ \hats ^2 + \hatu ^2 }{\hatt ^2}
+ \frac{ \hatt ^2 + \hats ^2 }{\hatu ^2}
-\frac{2}{3}
 \frac{\hats ^2 }{\hatt \hatu }\right]%\right)
~~~~ ~~~~\equiv QCD
\label{P+S17.71}
\eeq

\bea
 \frac{d \hat{\sigma}}{d \hatt}
(qq \to qq) &=&
\label{dsigdt4}
%\frac{1}{2}\left(
QCD -  \frac{ \eta_{8,XX} }{\Lambda^2  }  \frac{4 \pi  \alpha_s }{ 27 }
 \frac{\hats}{  \hatu \hatt } 
+
\frac{ |\eta_{8,XX}|^2 }{    \Lambda^4 }
\frac{4 \pi}{27}  %\frac{\hatt^2}{\hats}%\right)
~~~~~~~~{\cal O}^{8,RR}, {\cal O}^{8,LL}
\\ \label{35} &=& %\frac{1}{2}\left(
QCD + 
\label{dsigdt5} 
\frac{\eta_{8,XY}}{ \Lambda^2 }
   \frac{  4 \pi  \alpha_s} {9 \hats ^2 } 
  \frac{ \hatu^3 + \hatt^3}{\hatt \hatu  }   +
 \frac{ |\eta_{8,XY}|^2}{  \Lambda^4}
 \frac{2\pi }{ 9  } 
 \frac{ (\hatu^2 + \hatt^2)}{\hats ^2 }%\right)
~~~~~~~~{\cal O}^{8,RL} = {\cal O}^{8,RL}
\\
 &= & %\frac{1}{2}\left(
QCD-
\frac{  \eta_{1,XX} }{\Lambda^2}
\frac{4 \pi \alpha_s }{9 }
 \frac{\hats}{\hatu\hatt}
+
\frac{ |\eta_{1,XX}|^2}{ \Lambda^4}
\frac{ 4 \pi }{ 3}  %\frac{\hatt^2}{\hats} %\right)
\label{ubi2}
~~~~~~~~~~{\cal O}^{1 ,   LL}, 
{\cal O}^{1 , RR}  \\
& =& %\frac{1}{2}\left(
QCD+ \eta_{1,XY}
\frac{2 \pi \alpha_{em}  }{\Lambda^2}
\left(\frac{\hatu^2 }{\hatt }+ \frac{\hatt^2 }{ \hatu}\right)
 \frac{1}{\hats^2}
+
%  ~~~~~~~~~~ ~~~~~~~~~~
\frac{|\eta_{1,XY}|^2 }{ \Lambda^4}
 \pi 
\frac{(\hatt^2 + \hatu^2)}{ \hats^2}
\label{ubi31}%\right)
~~~~~~~~{\cal O}^{1,RL} = {\cal O}^{1,RL}
\label{37}
\eea
A factor of 1/2 for identical fermions in the final state
is included in eqs.~(\ref{P+S17.71}, \ref{ubi2}, and \ref{dsigdt4}). 
When the final
state quarks have opposite chirality, they are not identical,
 but  the operators ${\cal O}^{RL} = {\cal O}^{,RL}$ are, so 
its the same to include one operator for distinct fermions, or two
operators for identical fermions.

Interference with QED  is included for
the singlet  operators involving quarks of 
different chirality: ${\cal O}^{1,RL} + {\cal O}^{1,RL}$.
The  interference with QCD is absent because the
spinor traces vanish:  a non-zero trace must contain
an even number of {colour} matrices,
so both the QCD vertices must appear in the trace.
Therefore there is only one trace, in which
appear both contact vertices, with conflicting
chiral projection operators.  However, 
 QED has no {colour} matrices, so  there
is an interference term  with two spinor traces.

\subsubsection{$q q' \to q q'$ , $i = m  \neq j =n$. }

\bea
\frac{d \hat{\sigma}}{d \hatt}
(ud \to ud) & =&
    \frac{4 \pi \alpha_s^2 }{ 9 \hatt^2  } 
 \frac{ \hats ^2 + \hatu ^2}{ \hats ^2} ~~~~~~\equiv QCD' 
\label{P+S:17.64}  ~~~~~~~~~\\
 &=&
\label{dsigdt6a} 
QCD' +  
 \frac{\eta_{8,XX} }{  \Lambda^2 } 
 \frac{4 \pi \alpha_s  }{ 9 \hatt } 
+ 
\frac{|\eta_{8,XX}|^2}{  \Lambda^4}
 \frac{2 \pi  }{ 9  } % \frac{\hatt^2}{\hats}
~~~~~~~~~~{\cal O}^{8,XX }
 \\
 &=& \label{dsigdt6b} 
QCD' +   
 \frac{\eta_{8,XY} }{  \Lambda^2 }
\frac{4 \pi \alpha_s }{ 9  } 
 \frac{ \hatu ^2 }{ \hatt \hats ^2}
+
 \frac{|\eta_{8,XY}|^2}{  \Lambda^4} 
\frac{2 \pi  }{ 9  }
 \frac{ \hatu ^2}{ \hats ^2} ~~~~~~~~~~{\cal O}^{8,XY } \\
\\
 &=& \label{dsigdt6c} 
QCD' 
+2 \pi \eta_{1,XX}  \frac{\alpha_{em}}{\Lambda^2}
\frac{1}{\hatt}
%~~~~~~~~~ ~~~~~~~~~
+\pi 
 \frac{|\eta_{1,XX}|^2}{  \Lambda^4}
% \frac{\hatt^2}{\hats} 
 ~~~~~~~~~~{\cal O}^{1,RR }\\
 &=&
QCD' 
+2 \pi \eta_{1,LL}  \frac{\alpha_{em}}{\Lambda^2}
\frac{1}{\hatt} +
\frac{8 \pi}{9} \eta_{1,CC}  \frac{\alpha}{\Lambda^2}
\frac{1}{\hatt}
\nonumber \\
%~~~~~~~~~ ~~~~
&&~~~~~  \label{dsigdt6cCC} 
+\pi \left( 
 \frac{|\eta_{1,LL}|^2}{  \Lambda^4}
+\frac{2}{3} \frac{\eta_{1,CC} \eta_{1,LL}}{  \Lambda^2 \Lambda^2}
+ \frac{|\eta_{1,CC}|^2}{  \Lambda^4}
\right)% \frac{\hatt^2}{\hats}
  ~~~~~~~~~~{\cal O}^{1,LL } + {\cal O}^{1,CC }\\
&=& QCD'
+2 \pi \eta_{1,XY}  \frac{\alpha_{em}}{\Lambda^2}
\frac{ \hatu^2}{\hatt \hats^2} 
+\pi 
 \frac{|\eta_{1,XY}|^2}{  \Lambda^4}  \frac{ \hatu ^2}{ \hats ^2} 
~~~~~~~~~~{\cal O}^{1,XY } 
\label{dsigdt6d} \\
&=& QCD'  ~~~~~~~~~~~~~~~~
+
 \frac{\pi}{6  \Lambda^4}  \frac{ 4\hatu ^2 + 4\hatt^2 -\hats^2}{ \hats ^2} 
~~~~~~~~~~{\cal O}^{S1 } \label{OS1}\\
&=& QCD'  ~~~~~~~~~~~~~~~~
+
 \frac{\pi}{27  \Lambda^4}  \frac{2 \hatu ^2 + 2\hatt^2 +\hats^2}{ \hats ^2} 
~~~~~~~~~~{\cal O}^{S8 } \label{OS8}
\eea
Notice that for $ud \to ud$, ${\cal O}^{1,LR } $ is different
from ${\cal O}^{1,RL }$. so their contributions
are not summed in the above formulae.  

Interference with QED is included when there is no interference
with QCD, and the interference between  ${\cal O}^{1,LL }$
and ${\cal O}^{1,CC }$ is given, although we constrain the
two operators separately.

\subsubsection{$q' q' \to q  q $}

If the quark flavour changes at the contact interaction,
there is no interference with QCD.  However,
there are two contact interaction diagrams,
and an interference term  when the  initial
and final states contain  identical quarks. At
the LHC, this can describe $uu \to cc$,
$dd\to ss$, and $dd\to bb$.
\bea
 \frac{d \hat{\sigma}}{d \hatt}
&=&
~~~~~~~~~~ ~~~~~~~~~~
\frac{|\eta_{8,XX}|^2}{\Lambda^4 }
 \frac{4 \pi }{27  }
~~~~~~~~~~ {\cal O}^{8,XX } 
\label{ubi4di}
\\
& = &
~~~~~~~~~~ ~~~~~~~~~~
\frac{|\eta_{8,XY}|^2 }{ \Lambda^4}
 \frac{2\pi }{9 }
\frac{(\hatt^2 + \hatu^2)}{ \hats^2}
\label{ubi3di}
~~~~~~~~~~ {\cal O}^{8,XY } = {\cal O}^{8,YX } \\
&=& 
~~~~~~~~~~ ~~~~~~~~~~
\frac{|\eta_{1,XX}|^2}{\Lambda^4 }
 \frac{4 \pi }{3 }  
~~~~~~~~~~{\cal O}^{1,XX } 
\label{ubi2bdi}
\\
& = &   
~~~~~~~~~~ ~~~~~~~~~~
\frac{|\eta_{1,XY}|^2}{\Lambda^4 }
\pi 
\frac{(\hatt^2 + \hatu^2)}{ \hats^2}
\label{ubi3bdi}
~~~~~~~~~~ {\cal O}^{1,XY }  =  {\cal O}^{1,YX } 
\eea
A factor of 1/2 for identical fermions in the final state
is included. In practice,  this list is just last terms
from  $qq \to qq$.

\subsubsection{$q' q^{''} \to q  q $, or
$q q \to q'  q^{''} $}

In the case where there are identical fermions
in either the initial or final states, but not both
(at the LHC: $uu \to uc$, $dd\to ds$,  $dd\to db$,
 $dd\to sb$),
there are still two diagrams, but no interference term:
\bea
\frac{d \hat{\sigma}}{d \hatt}
(dd \to  s  b)
&=&
~~~~~~~~~~ ~~~~~~~~~~
\frac{|\eta_{8,XX}|^2}{\Lambda^4 }
 \frac{4 \pi }{9  }  
~~~~~~~~~~ {\cal O}^{8,XX } 
\label{ubi4}
\\
& = &
~~~~~~~~~~ ~~~~~~~~~~
\frac{|\eta_{8,XY}|^2 }{ \Lambda^4}
 \frac{2 \pi }{9 }
\frac{(\hatt^2 + \hatu^2)}{ \hats^2}
\label{ubi3}
~~~~~~~~~~ {\cal O}^{8,XY } \\
&=& 
~~~~~~~~~~ ~~~~~~~~~~
\frac{|\eta_{1,XX}|^2}{\Lambda^4 }
2 \pi  
~~~~~~~~~~{\cal O}^{1,XX } 
\label{ubi2b}
\\
& = &   
~~~~~~~~~~ ~~~~~~~~~~
\frac{|\eta_{1,XY}|^2}{\Lambda^4 }
\pi 
\frac{\hatt^2 + \hatu^2}{ \hats^2}
\label{ubi3b}
~~~~~~~~~~ {\cal O}^{1,XY } 
\eea
and in the case where the identical fermions are in
the final state, the given formulae should be multiplied by
1/2.

\subsubsection{$q^{''} q \to q q'$ 
and any vertex with more
than three different flavours}

At the LHC this can describe $ud\to cs$, 
$ud\to cb$,  and also $ud\to us$, $ud\to ub$,
$ud\to cd$:
\bea
\frac{d \hat{\sigma}}{d \hatt}
& =& \label{dsigdt1a}
~~~~~~~~~~ ~~~~~~~~
\frac{|\eta_{8,XX}|^2}{  \Lambda^4}
\frac{2 \pi }{ 9 }  
~~~, ~~~
{\cal O}^{8,XX } \\
& =& 
~~~~~~~~~~ ~~~~~~~~
\frac{|\eta_{8,XY}|^2}{  \Lambda^4}
\frac{2 \pi }{ 9 }
\frac{ \hatu ^2}{\hats ^2}  
\label{dsigdt1b}
~~~, ~~~
{\cal O}^{8,XY }
\\ 
& = & \label{dsigdt1c}
~~~~~~~~~~ ~~~~~~~~
\frac{|\eta_{1,XX}|^2}{  \Lambda^4}
\pi  
~~~, ~~~
{\cal O}^{1,XX } ,{\cal O}^{1,CC } \\
& =& \label{dsigdt1d}
~~~~~~~~~~ ~~~~~~~~
\frac{|\eta_{1,XY}|^2}{  \Lambda^4}
\pi 
\frac{ \hatu ^2}{\hats ^2} 
~~~, ~~~
{\cal O}^{1,XY }\\
& =& \label{OS1FC}
~~~~~~~~~~ ~~~~~~~~
 \frac{\pi}{12  \Lambda^4}  \frac{ 4\hatu ^2 + 4\hatt^2 -\hats^2}{ \hats ^2} 
~~~~~~~~~~{\cal O}^{S1 }\\
& =& \label{OS8FC}
~~~~~~~~~~ ~~~~~~~~
 \frac{\pi}{54 \Lambda^4}  \frac{2 \hatu ^2 + 2\hatt^2 +\hats^2}{ \hats ^2} 
~~~~~~~~~~{\cal O}^{S8 } 
\eea

\subsection{$q_i \bar{q}_m \to \bar{q}_j q_n$}

For contact interactions with  two incident  
first generation quarks, the best bounds arise from
$qq \to qq$. However,  a ``flavour diagonal''
interaction involving a  quark and anti-quark
of the first generation, going to  a quark and anti-quark
of a higher generation, is better constrained by
the Tevatron, who had  valence $\overline{q}_1 q_1$  
in the initial state.  The
cross sections for  contact interactions in
quark-anti-quark collisions, can be
obtained by crossing  (\ref{SDcrossingrules}),
the previous formulae, and removing,
if neccessary, the factor of 1/2 for identical
fermions in the final state.

\end{document}